\documentclass[preprint,showpacs,amsmath,amssymb,pdflatex]{revtex4}
\pdfoutput=1
\usepackage{graphicx}
\usepackage{dcolumn}
\usepackage{bm}
\begin{document}

\title{Green's functions for solving differential equations,\\ in non-boundary
value problems in near-field optics\\ and in quantum transport through point
contacts}

\author{Ursula Schr\"oter}
\email{Ursula.Schroeter@uni-konstanz.de}
\affiliation{Fachbereich Physik, Universit\"at Konstanz,
Universit\"atsstra\ss e 10, 78457 Konstanz, Germany}

\date{\today}

\begin{abstract}
This introduction to Green's functions is based on their
role as kernels of differential equations. The procedures to construct
solutions to a differential equation with an external source or with an
inhomogeneity term are put together to derive the Dyson equation for the
Green's function of the inhomogeneous system. Very different areas of physics
such as, for example, electrodynamics and quantum transport, can profit from
this Green's function formalism.

The fundamental homogeneous-medium Green's tensor of electrodynamics is
deduced from the field of a dipole. Based upon that a numerical procedure
is presented to solve the wave-equation for the near-field in a scattering
setup for arbitrary material distributions. The full inhomogeneous system's
Green's function is not explicitly needed to get the fields, although it can
be obtained by a very similar calculation and in optics can be interpreted as
a density of states. 

It is demonstrated how the transport problem for two open free-electron gas
reservoirs with arbitrary coupling can be solved by finding the system's
Green's function. In this sense the article is an introduction on Green's
functions for treating interaction.
A very detailed discussion of the current formula is given on an elementary
basis.
\end{abstract}


\maketitle

\renewcommand\thesubsection {\thesection.\arabic{subsection}}

\section{Green's functions tool for solving differential equations}

Green's functions \cite{Ohtaka,Don,Koo,Frie}
are encountered as response functions, time-ordered
expectation values, certain solutions of boundary-value problems or
resolvent kernels. This introduction to Green's functions is based on their
role as kernels of differential equations. The procedures to construct
solutions to a differential equation with an external source or with an
inhomogeneity term are put together to {\it derive} the Dyson equation for the
Green's function of the inhomogeneous system. Very different areas of physics
such as, for example, electrodynamics (see section II and \cite{Chic})
and quantum transport (see section III and \cite{Sche98,Cue}), can profit from
such Green's function formalisms.

\subsection{Introduction}
Green's function formalisms do not present a cure-all for solving differential
equations, because essentially the problem of finding the solution of the
differential equation is shifted to that of finding the corresponding Green's
function. This can, however, be a simplification and even give access to the
solution of a more general class of problems. Starting from ordinary
differential equations, this short review is meant to introduce how the
corresponding Green's functions are defined and how they are involved in
constructing the solutions for different types of differential equations.
Although bearing similarities, the Green's function formalism can go
beyond perturbation theory. Special emphasize is laid here upon the parallels
between homogeneous and inhomogeneous ordinary differential and Green's
functions equations.

\subsection{Homogeneous equation}
The starting point is a homogeneous differential equation
\begin{equation}
D_{\xi}\; \phi_0(\xi) = 0
\end{equation}
which we suppose is exactly solvable, although $\phi_0(\xi)$ will not
explicitly be needed. $D_{\xi}$ is some differential operator which may include
multiplication with a constant or even another function of $\xi$. $\xi$ is
either a space or time variable. What we need to know is the solution $g$
of the corresponding Green's functions equation
\begin{equation}
D_{\xi}\; g(\xi,\xi') = \delta(\xi-\xi')
\end{equation}
$g$ is a tensor-like function of two arguments, $D_{\xi}$ only acting on
the first of them. Like $g$ replaces $\phi$ on the left, the zero on the
right side of (1) is replaced by a $\delta$-distribution in (2). There is no
general recipe, but knowing $\phi_0(\xi)$ can help to get $g(\xi,\xi')$. 

\subsection{Source term}
Having $g$, the construction of a solution
\begin{equation}
\phi_q(\xi)=\phi_0(\xi)+\int d\xi' \; g(\xi,\xi') \; Q(\xi')
\end{equation}
of the differential equation with a source term $Q$ on the right side
\begin{equation}
D_{\xi} \; \phi_q(\xi) = Q(\xi)
\end{equation}
is straight forward. Of course, a solution $\phi_0$ of (1) can be added
independently of $Q$, so we only need to proove that the integral term from (3)
satisfies (4):
\begin{eqnarray}
D_{\xi} \int d\xi' \; g(\xi,\xi') \; Q(\xi') &=& 
\int d\xi' \; [D_{\xi} \; g(\xi,\xi')] \; Q(\xi') \nonumber \\ &=&
\int d\xi' \; \delta(\xi-\xi') \; Q(\xi') = Q(\xi) \nonumber 
\end{eqnarray}
(All integrals are understood to range over the entire $\xi$-space.)

\subsection{Inhomogeneity}
Instead of a source term, the differential equation can contain a potential
term which we shall call an inhomogeneity.
\begin{equation}
D_{\xi} \; \phi_{ih}(\xi) - V(\xi) \; \phi_{ih}(\xi) = 0
\end{equation}
that is a $\xi$-dependent function $V$ multiplied with $\phi$. $V(\xi)$ is
not included in $D_{\xi}$, because we assume that it so much complicates the
equation that a standard solution is no longer known. The minus sign is a
useful convention. We put the inhomogeneity term on the right
\renewcommand{\theequation}{\arabic{equation}a}
\setcounter{equation}{4}
\begin{equation}
D_{\xi} \; \phi_{ih}(\xi) = V(\xi) \; \phi_{ih}(\xi) 
\end{equation}
\renewcommand{\theequation}{\arabic{equation}}
to make (5) formally look like (4). With $V\phi$ playing the role of $Q$ a
formal solution is constructed analogously to (3):
\begin{equation}
\phi_{ih}(\xi) = \phi_0(\xi) + \int d\xi' \; g(\xi,\xi') \; V(\xi') \; \phi_{ih}(\xi')
\end{equation}
This presents an implicit equation for $\phi_{ih}$, the so-called
{\it Lippmann-Schwinger} equation. We have replaced the differential
equation by an integral equation. Inserting the solution of the homogeneous
equation $\phi_0$ also for $\phi_{ih}$ in the integral on the right side of (6)
would give the {\it Born} approximation $\phi_{ih}(\xi)\approx \phi_0(\xi)+
\int d\xi' \; g(\xi,\xi') \; V(\xi') \; \phi_0(\xi')$,
which is appropriate if $V\phi$ is
a small perturbation compared to $D_{\xi}\phi$. However, it is the virtue of
the Green's functions method that in contrast to perturbation theory the
inhomogeneity need {\it not} be a small deviation. Although not yet providing
an explicit solution for $\phi_{ih}$, one can make use of
(6) in numerical calculations (see section II and \cite{DerGir}).
As (2) is the corresponding Green's function equation with a point source to
the homogeneous equation (1), the equation defining the Green's function $G$
for the inhomogeneous case derived from (5) reads:
\begin{equation}
[D_{\xi}-V(\xi)] \; G(\xi,\xi') = \delta(\xi-\xi') 
\end{equation}
However, while supposing that we know or can easily guess $g(\xi,\xi')$,
there is no hint yet how to calculate $G(\xi,\xi')$.

\subsection{General case}
Finally we have to treat the most general case with source and inhomogeneity
term:
\begin{equation}
D_{\xi} \; \phi_{ihq}(\xi) - V(\xi) \; \phi_{ihq}(\xi) = Q(\xi)
\end{equation}
In analogy to (3) and (4) a yet formal solution can be written down using (7):
\begin{equation}
\phi_{ihq}(\xi) = \phi_{ih}(\xi) + \int d\xi' \; G(\xi,\xi') \; Q(\xi')
\end{equation}
Here we have the freedom to add any solution of (6) to the integral.

But let us start again solving the problem directly from (8) which we rewrite
as
\renewcommand{\theequation}{\arabic{equation}a}
\setcounter{equation}{7}
\begin{equation}
D_{\xi} \; \phi_{ihq}(\xi) - Q(\xi) = V(\xi) \; \phi_{ihq}(\xi) 
\end{equation}
\renewcommand{\theequation}{\arabic{equation}}
\setcounter{equation}{9}
We compare this with (5a) and construct a solution of the same form as (6).
$\phi_q$ takes the role of $\phi_0$, because it would satisfy (8a) if the
right side were zero. $\phi_{ih}$ has to be replaced by $\phi_{ihq}$. Using
(3) for $\phi_q$ we obtain
\begin{equation}
\phi_{ihq}(\xi) = \underbrace{ \phi_0(\xi) + \int d\xi' \; g(\xi,\xi') \;
Q(\xi') }_{\phi_q} + \int d\xi'' \; g(\xi,\xi'') \; V(\xi'') \;
\phi_{ihq}(\xi'')
\end{equation}
(10) could also have been established by a different line of thought. In (8)
putting everything that differs from the homogeneous equation (1) on the
right side gives
\renewcommand{\theequation}{\arabic{equation}a}
\setcounter{equation}{7}
\begin{equation}
D_{\xi} \; \phi_{ihq} = Q(\xi) + V(\xi) \; \phi_{ihq}(\xi)
\end{equation}
\renewcommand{\theequation}{\arabic{equation}}
\setcounter{equation}{10}
If it were not for the inhomegeneity term $V\phi$, the solution would be (3).
And if it were not for the source term $Q$, we could use (6). Although this is,
of course, not a correct way to solve non-homogeneous differential equations,
we can understand (10) as an ansatz adding these two contributions. The
$\phi_0$-part does not have to be written twice. And $\phi_{ihq}$ has to
appear instead of $\phi_{ih}$ in the integral taken from (6). This still
unknown $\phi_{ihq}$ leaves the neccessary freedom to somehow counterbalance
the $Q$-term
not present in (5) and (6), which justifies (10) as an ansatz.
Whichever way it was obtained, (10) is an implicit integral equation for
$\phi_{ihq}$ as (6) is for $\phi_{ih}$. 

\subsection{Dyson equation}
Now we shall profit from the fact
that with (9) we have a second representation of $\phi_{ihq}$. Insert the
expression from (9) for $\phi_{ihq}$ both on the left and on the right side in
(10):
\begin{eqnarray} 
\phi_{ih}(\xi) &+& \int d\xi' \; G(\xi,\xi') \; Q(\xi') = 
\phi_0(\xi) + \int d\xi' \; g(\xi,\xi') \; Q(\xi') + \nonumber \\ &&
\int d\xi'' g(\xi,\xi'') \; V(\xi'') \;
[\; \phi_{ih}(\xi'') + \int d\xi' \; G(\xi'',\xi') \; Q(\xi') \; ]
\nonumber \end{eqnarray}
$\phi_{ih}$ on the left cancels with $\phi_0$ and one integral term on the
right according to (6) and we are left with
\begin{eqnarray}
\int d\xi' \; G(\xi,\xi') \; Q(\xi') &=& \int d\xi' \; g(\xi,\xi') \; Q(\xi')
+ \nonumber \\ &&
\int d\xi'' \int d\xi' \; g(\xi,\xi'') \; V(\xi'') \; G(\xi'',\xi') \; Q(\xi')
\nonumber \end{eqnarray}
This is an implicit equation for the unknown Green's function $G$. But as $G$
by definition (7) does not depend on any source term, (11) must be valid for
arbitrary $Q$. The special choice $Q(\xi')=\delta(\xi'-\xi_0)$ gives
\begin{eqnarray}
\int d\xi' \; G(\xi,\xi') \; \delta(\xi'-\xi_0) &=& \int d\xi' \; g(\xi,\xi') \;
\delta(\xi'-\xi_0) + \nonumber \\ &&
\int d\xi'' \int d\xi' \; g(\xi,\xi'') \; V(\xi'') \;
G(\xi'',\xi') \; \delta(\xi'-\xi_0)
\nonumber \end{eqnarray}
which after carrying out the $\xi'$-integrals and then renaming $\xi_0$ to
$\xi'$ becomes
\begin{equation}
G(\xi,\xi') = g(\xi,\xi') + \int d\xi' \; g(\xi,\xi'') \; V(\xi'') \;
G(\xi'',\xi')
\end{equation}
(11) is the general implicit integral equation for the inhomogeneous system's
Green's function $G$ and called {\it Dyson} equation.

We could have derived (11) by doing all calculations from (8) on with only a
point source $\delta(\xi-\xi_0)$ instead of $Q(\xi)$. Putting in the solutions
(3) and (9) for this case would just have looked a bit awkward.
(11) is not an explicit solution for $G$, but still an implicit equation,
and even though it is often written as $G=g+gVG$ one must not forget that
there is a convolution-like integration over the inner $\xi$-index behind
the sequence of factors $gVG$. The
integration can either be transformed into a discrete finite sum, and thus (11) into
a linear system of equations solvable by a matrix inversion
(see section II and \cite{EPJ}), or the
convolution can be replaced by a multiplication going to frequency space by
a Fourier transformation (see section III and \cite{Cue}).

(10) and the preparation for (11) would have looked much more elegant leaving
out $\phi_0$ and the $\phi_{ih}$-terms right from the start. Indeed, one could
argue that in (3) one is only interested in the part different from the
trivial homogenous solution, if no further boundary conditions have to be
accounted for. With a similar argument, the $\phi_{ih}$-part could have been
dropped in (9). The identities
\renewcommand{\theequation}{\arabic{equation}*}
\setcounter{equation}{2}
\begin{equation}
\phi_q^*(\xi) = \int d\xi' \; g(\xi,\xi') \; Q(\xi')
\end{equation}
\renewcommand{\theequation}{\arabic{equation}*}
\setcounter{equation}{5}
\begin{equation}
\phi_{ih}^*(\xi) = \int d\xi' \; g(\xi,\xi') \; V(\xi') \; \phi_{ih}^*(\xi')
\end{equation}
\renewcommand{\theequation}{\arabic{equation}*}
\setcounter{equation}{8}
\begin{equation}
\phi_{ihq}^*(\xi) = \int d\xi' \; G(\xi,\xi') \; Q(\xi')
\end{equation}
\renewcommand{\theequation}{\arabic{equation}*}
\setcounter{equation}{9}
\begin{equation}
\phi_{ihq}^*(\xi) = \int d\xi' \; g(\xi,\xi') \; Q(\xi') +
\int d\xi'' \; g(\xi,\xi'') \; V(\xi'') \; \phi_{ihq}^*(\xi'') 
\end{equation}
\renewcommand{\theequation}{\arabic{equation}}
\setcounter{equation}{11}
would obviously also have given (11). Leaving out $\phi_0$ in (6) would be
precarious (see below). However, dropping it when
constructing (11) along our second line of thought causes no problem.
The term with $Q$ will ensure that $\phi_{ihq}^*$ cannot simply be zero.

\subsection{Expansions}
There is some subtlety about the $\phi_0$ contribution in (6).
(6*) has the trivial solution $\phi_{ih}^*(\xi)=0$ and not necessarily
another one. Surely, $\phi_{ih}(\xi)=0$ satisfies (5), but it is not what
we want. $\phi_0$ is the background excitation replacing boundary conditions
in this kind of problem, which shall become clear when discussing
electrodynamics (see following article). $\phi_0$ acts as a source term which -
instead of being put in $Q$ - can be more appropriately and conveniently set
as a fixed part of the $\phi$ we are looking for.
Inserting (6) into itself ever
and ever again, $\phi_{ih}$ is developed into a series in powers of $gV$:
\begin{eqnarray}
\phi_{ih} &=& \phi_0 + \int gV\phi_{ih}  \nonumber \\ &=&
\phi_0 + \int gV (\phi_0 + \int gV (\phi_0 + \int gV ( \ldots \; \ldots \;
\ldots ))) \nonumber \\ &=& \phi_0 + gV\phi_0 + gVgV\phi_0 + \ldots =
\sum_{n=0}^{\infty} (gV)^n \phi_0
\end{eqnarray}
Just for shorthand notation we dropped the arguments and integration variables
and in the third line also the integral signs. The resulting series is
recognized from perturbation theory summing
interactions to zeroth, first, second, etc. order. If there were no
contribution $\phi_0$ in $\phi_{ih}$, only the contribution of $gV$ to
power infinity would exist with nothing to multiply to at the end. In other
words we would have no basis on which to develop $\phi_{ih}$. The
Green's functions formalism as we use it aims at solving (6), (10) or
(11) in a closed form, {\it not} cutting the series in (12) at some finite
order.
The development (12) is only shown here precisely to demonstrate that
interactions are included to all orders as well as to explain the importance
of the homogeneous background contribution.
The Dyson equation (11) can be expanded in an analogous manner to the
Lippmann-Schwinger equation, by the way prooving the equivalence to its complementary
form $G=g+GVg$:
\begin{equation}
G = g + \int gVG = g \sum_{n=0}^{\infty} (Vg)^n 
= \sum_{n=0}^{\infty} (gV)^n \; g = g + \int GVg
\end{equation}
(Integration over inner arguments is understood in all contributions to
the sums.)
Either formally or as a matrix calculation in finite discrete $\xi$-space,
(11) is often solved as
\begin{equation}
G=({\bf 1}-gV)^{-1}g
\end{equation}
Developing the $(\ldots)^{-1}$-factor in (14) into a geometric series just
results in the infinite sums written in (13).

Putting together (12) and (13) we can get an alternative representation to
(6) for the solution of the inhomogeneous differential equation:
\begin{equation}
\phi_{ih}(\xi)=\phi_0(\xi)+\int d\xi' \; G(\xi,\xi') \; V(\xi') \; \phi_0(\xi')
\end{equation}
Therefore in the case that should it be easier to get the Green's function $G$
than to solve the implicit equation (6) for $\phi_{ih}$, we see that $G$ can
be useful also for treating the equation without an external source $Q$.
Nevertheless, (15) again illustrates that $V\phi_0$ plays the role of the
source.

\subsection{Conclusions and Outlook for section I}
Using Green's functions it has been shown how differential equations can be
treated that differ from easily solvable ones by an additional potential
term or an arbitrary source term. The formalism as presented here is for
open-boundary in contrast to boundary-value problems \cite{Koo,Jac}.
From constructing
the solutions of the differential equations, we also obtained the constituting
relation for the inhomogeneous system's Green's function, which characterizes
the response to a point source and includes all-order interactions.
Examples of applications will be given in the two following sections.

\newpage
\setcounter{equation}{0}

\section{Non-boundary value problems in near-field optics}

The fundamental homogeneous-medium Green's tensor of electrodynamics is
deduced from the field of a dipole. Based upon that a numerical procedure
is presented to solve the wave-equation for the near-field in a scattering
setup for arbitrary material distributions. The full inhomogeneous system's
Green's function is not explicitly needed to get the fields, although it can
be obtained by a very similar calculation and in optics can be interpreted as
a density of states. 

\subsection{Introduction}
The typical problem in nano-optics \cite{Gre97,DerChem} is the situation that some tiny structures
are illuminated by an extended source, a plane wave for example, and then
one is interested in the field distribution that arises from multiple
scattering \cite{scat}, especially to identify places where the field intensity gets
considerably enhanced \cite{Martinres,Ebbfluo}.
A theory can be based on Green's functions, however, their implication differs
slightly from the standardly taught cases of fixed boundary field values
\cite{Jac_1} or
located sources. Modern optical scanning microscopes make it possible to
probe and map directly even different quantities of the near-field
\cite{interpret}, such as
the electric and magnetic field intensities \cite{Eloise}
or the density of states \cite{Chicel}.
Applications of tayloring nano-structures with respect to optical properties
include resonant particles \cite{dicro,partweb} and cavities
\cite {Krenncav,Perney}, squeezed fields \cite{squeeze},
wave guides and their adressing \cite{adres,MarLith,Quidwave}
as well as transmission apertures \cite{Perney,Thio} and lithography masks
\cite{MarLith}.
We here present the Green's functions formalism that forms the bases of a
finite-element quite effective numerical algorithm used in current research
\cite{girneu,EPJel,coral}.
This treatise is also given as an application example of the general methods
to solve differential equations with certain perturbations presented in the
preceeding paper.

\subsection{Problem}

The discussion can be reduced to monochromatic light, that is a single
frequency $\omega$ and thus time dependence $e^{-i\omega t}$ for the fields.
With non-magnetic materials the problem is to find the solution 
$\vec{E}(\vec{r})$ of the wave equation
\begin{equation}
-\vec{\nabla} \times \vec{\nabla} \times \vec{E}(\vec{r}) + \frac{\omega^2}{
c^2} \; \varepsilon(\vec{r}) \; \vec{E}(\vec{r})=0
\end{equation}
(1) is a differential equation of the type (5) from section I.
There is no source term on the right hand side of (1). If the source were,
for example, a dipole located at some point, a source term with its oscillation
strength and direction would have to be put on the right as
$\vec{Q}\delta(\vec{r}-\vec{r}_0)$.
However, we shall see that a plane-wave source can be and is better included
in (1) as it is. For simplicity we shall assume that the background medium,
in which objects with different permittivities $\varepsilon$ are located (Fig.1),
is vacuum with permittivity $\varepsilon_0$. For another embedding medium, its
dielectric constant $\varepsilon_b$ would take the role of $\varepsilon_0$.
In (1) $\varepsilon(\vec{r})$ of the material distribution designates the
dimensionless relative permittivity with respect to vacuum or the background
medium. To separate (1) into a part representing a homogeneous differential
equation with known solution and an inhomogeneity write its as
\renewcommand{\theequation}{\arabic{equation}a}
\setcounter{equation}{0}
\begin{equation}
-\vec{\nabla} \times \vec{\nabla} \times \vec{E}(\vec{r}) + \frac{\omega^2}{
c^2} \; \vec{E}(\vec{r}) + \frac{\omega^2}{c^2} \; (\varepsilon(\vec{r})-1) \;
\vec{E}(\vec{r}) = 0
\end{equation}
\renewcommand{\theequation}{\arabic{equation}}
The correspondances to the quantities of the general formalism given in
section I are
$$ \xi \leftrightarrow \vec{r}, \;
D_{\xi} \ldots \leftrightarrow -\vec{\nabla} \times \vec{\nabla} \times
\ldots+\frac{\omega^2}{c^2}\cdot \ldots, \;
V \leftrightarrow -\frac{\omega^2}{c^2}(\varepsilon-1), \;
\phi_{ih} \leftrightarrow \vec{E}$$
and $\phi_0$ will become $\vec{E}_b$. We can write down the solution
following section I after having prepared the background Green's function in the
next section.

\begin{figure}
\includegraphics[width=10cm,angle=270]{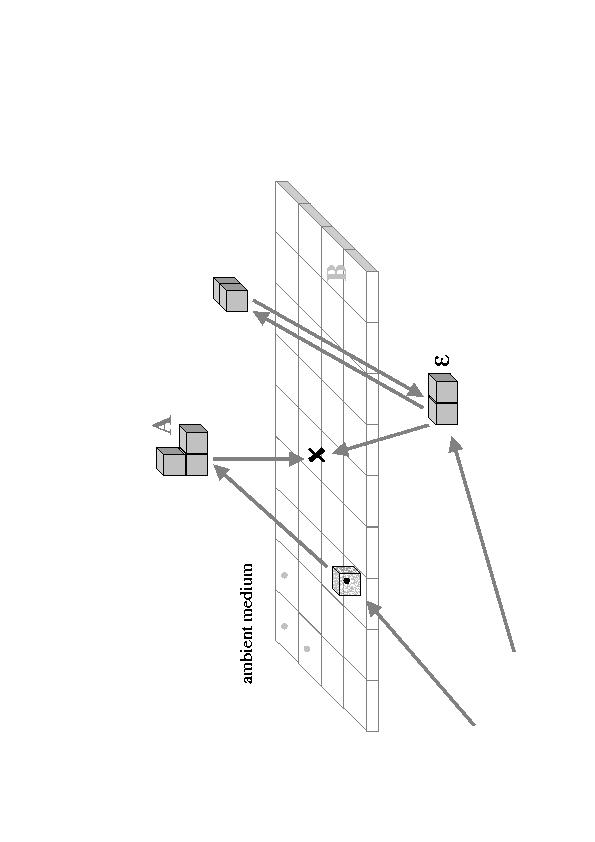}
\caption{Objects distributed in space and discretized into cubes of
equal size. A plane in space discretized into a quadratic mesh is also shown.
Arrows indicate only some possible scattering paths to a location marked by
the cross.}
\end{figure}

\subsection{Background Green's function}
In any case we need the Green's function $g(\vec{r},\vec{r'})$ of the
homogeneous problem satisfying
\begin{equation}
-\vec{\nabla}_{\vec{r}} \times \vec{\nabla}_{\vec{r}} \times g(\vec{r},\vec{r'})
+k^2g(\vec{r},\vec{r'})={\bf 1} \; \delta(\vec{r}-\vec{r'})
\end{equation}
with $k=\omega /c$.
$g$ will be a $3\otimes 3$ tensor or matrix here. More commonly \cite{Chew}
the small letter $g$ is used for the scalar function
\begin{eqnarray}
g_{scal}(\vec{r},\vec{r'})=-\frac{e^{ik\vert \vec{r}-\vec{r'} \vert}}{
4\pi \vert \vec{r}-\vec{r'} \vert} \quad {\rm fulfilling} \nonumber \\
\triangle_{\vec{r}} \; g_{scal}(\vec{r},\vec{r'})+k^2 g_{scal}(\vec{r},\vec{r'})
=\delta(\vec{r}-\vec{r'}) \end{eqnarray}
with $k=\omega /c$. The Green's function $g$ from (2) is then named $G^h$
with index $h$ for {\it homogeneous}. There are several ways to obtain $g$.
One is based on the knowledge that if we have a scalar function $\Psi(\vec{r})$
solving $\triangle\Psi+k^2\; \Psi=0$, then $\vec{F}_1=\vec{\nabla} \times
(\vec{a}\Psi)$ and $\vec{F}_2=\vec{\nabla} \times \vec{\nabla} \times (\vec{
a}\Psi)$ with a constant but arbitrary pivot vector $\vec{a}$ will both solve
the vectorial equation $-\vec{\nabla} \times \vec{\nabla} \times \vec{F} +
k^2 \vec{F}=0$. The tensor $g$ looked for in (2) can be constructed out of
$\vec{F}_1(\vec{r})$, $\vec{F}_1(\vec{r'})$, $\vec{F}_2(\vec{r})$,
$\vec{F}_2(\vec{r'})$ together with $\vec{F}_3(\vec{r})=\vec{\nabla}\Psi(\vec{r})$
and $\vec{F}_3(\vec{r'})$.
We shall not enter into the details of this mathematically slightly precarious
approach \cite{AD}. A second recipe just mentioned here for completeness is
given by the following statement \cite{Chew}:
If $g_{scal}(\vec{r},\vec{r'})$ satisfies (3), then
\begin{equation}
g(\vec{r},\vec{r'})=({\bf 1}+\frac{\vec{\nabla}_{\vec{r}} \otimes
\vec{\nabla}_{\vec{r}}}{k^2}) \; g_{scal}(\vec{r},\vec{r'})
\end{equation}
is the tensor defined by (2). Of course, because of being in homogeneous space
$g_{scal}$ and $g$ effectively are functions of $\vec{R}=\vec{r}-\vec{r'}$
alone. ${\bf 1}$ is the unit matrix in 3 by 3 cartesian coordinate space and
\newline
$\vec{\nabla} \otimes \vec{\nabla}$ means building a matrix out of derivatives
$\begin{pmatrix}
\partial^2_x & \partial_x \partial_y & \partial_x \partial_z \cr
\partial_y \partial_x & \partial^2_y & \partial_y \partial_z \cr 
\partial_z \partial_x & \partial_z \partial_y & \partial^2_z \cr\end{pmatrix}$. 

We shall deduce $g$ from a physical reasoning.
From standard electrodynamics \cite{Jac_2}
one has the electric field of an oscillating dipole
\begin{equation}
\vec{E}(\vec{R})=\frac{k^2e^{ikR}}{4\pi\varepsilon_0R}\left(
\frac{\vec{p}R^2-\vec{R}(\vec{R}\vec{p})}{R^2}+\frac{3\vec{R}(\vec{R}\vec{p})
-\vec{p}R^2}{R^2}\left(\frac{1}{k^2R^2}-\frac{i}{kR}\right)\right)
\end{equation}
It is important to take the exact formula here including retardation in
contrast to common near- or far-field approximations. (5) gives the space
part, the time dependence is just $e^{-i\omega t}$ everywhere. To get the
Green's tensor, we have to evaluate from (5) what field components in
x-, y- and z-direction a dipole $\vec{p}$ at $\vec{r'}$ oriented along x
would produce at $\vec{r}$, what components a dipole oriented along y would
produce and what components a dipole along z would produce and assemble all
these in a matrix. A point dipole is the elementary excitation corresponding
to the $\delta$ on the right side of (2). The physical meaning of $g$ is to
tell us what field any such dipole would have. That is shown formally in
the first matrix in (5a). Decomposing any $\vec{p}$ into its cartesian
components, (5) can be rewritten as
$$\vec{E}(\vec{R})=\begin{pmatrix}
E_x(\vec{r}) \leftarrow p_x(\vec{r'}) & \phantom{e} &
E_x(\vec{r}) \leftarrow p_y(\vec{r'}) & \phantom{e} &
E_x(\vec{r}) \leftarrow p_z(\vec{r'}) \cr
E_y(\vec{r}) \leftarrow p_x(\vec{r'}) & \phantom{e} &
E_y(\vec{r}) \leftarrow p_y(\vec{r'}) & \phantom{e} &
E_y(\vec{r}) \leftarrow p_z(\vec{r'}) \cr
E_z(\vec{r}) \leftarrow p_x(\vec{r'}) & \phantom{e} &
E_z(\vec{r}) \leftarrow p_y(\vec{r'}) & \phantom{e} &
E_z(\vec{r}) \leftarrow p_z(\vec{r'}) \cr
\end{pmatrix} \begin{pmatrix} p_x \cr p_y \cr p_z \cr \end{pmatrix}$$

\begin{center} ${\displaystyle =
\frac{k^2e^{ikR}}{4\pi\varepsilon_0R}\left( \left( 1-\frac{1-ikR}{k^2R^2} \right)
\begin{pmatrix} 1 & 0 & 0 \cr 0 & 1 & 0 \cr 0 & 0 & 1 \cr \end{pmatrix}
\begin{pmatrix} p_x \cr
p_y \cr p_z \cr \end{pmatrix} - \frac{-3+3ikR+k^2R^2}{
k^2R^4} \begin{pmatrix} X^2 & XY & XZ \cr YX & Y^2 & YZ \cr ZX & ZY & Z^2 \cr \end{pmatrix}
\begin{pmatrix} p_x \cr p_y \cr p_z \cr \end{pmatrix} \right)
}$ \end{center}

$$ \eqno (5a) $$

from which we easily see that the matrix to be multiplied with $\vec{p}$
to produce $\vec{E}(\vec{R})$ is
$$
\frac{k^2e^{ikR}}{4\pi\epsilon_0R}\left( {\bf 1} \; \left( 1-\frac{1-ikR}{
k^2R^2}\right) - \vec{R} \otimes \vec{R} \; \frac{-3+3ikR+k^2R^2}{k^2R^4} \right)
$$
$R$ without arrow means the absolut value $\vert \vec{R} \vert = \sqrt{
X^2+Y^2+Z^2}$ and $X$, $Y$, $Z$ stand for $x-x'$, $y-y'$ and $z-z'$,
respectively. $\vec{R}\otimes \vec{R}$ is the matrix written with $X$, $Y$
and $Z$ from (5a). The terms from (5) vectorially oriented along $\vec{p}$
cause the diagonal matrix contribution, those stemming from terms
with $\vec{R}(\vec{R}\vec{p})$ the full matrix in (5a). Compared to the
above expression $g(\vec{R})$ from (4) has a minus sign and misses a factor
$k^2/\varepsilon_0$. As will be discussed later, the source to put into
equation (1) corresponding to an oscillating dipole is not the dipole moment
$\vec{p}$ itself, but $-\mu_0\omega^2$ times $\vec{p}$. And because
$-\frac{k^2}{\varepsilon_0}\frac{1}{\mu_0\omega^2}=-1$ we have
\begin{equation} g(\vec{R})=
-\frac{e^{ikR}}{4\pi R}\left( {\bf 1} \; \left( 1-\frac{1-ikR}{
k^2R^2}\right) - \vec{R} \otimes \vec{R} \; \frac{-3+3ikR+k^2R^2}{k^2R^4} \right)
\end{equation}
The formula (6) fails for $\vec{r}=\vec{r'}$ or $\vec{R}=0$. $g(\vec{R})$
including the case $\vec{R}=0$ can be represented using the principal volume
method \cite{Chew}. In practice, working with finite elements, the value to put for
$g$ of its two arguments the same place can be derived from the polarization
of a dielectric body. The discussion of $g(\vec{r},\vec{r})$ is postponed
to the next section.

\subsection{Solution for the field}
The starting point to find a solution for the electric field with the objects
present is eq. (6) of section I, which rewritten in the variables of our
problem here reads
\begin{equation}
\vec{E}(\vec{r})=\vec{E}_b(\vec{r})+\int d^3\vec{r'} \; g(\vec{r},\vec{r'}) \;
V(\vec{r'}) \;\vec{E}(\vec{r'}) 
\end{equation}
where $\vec{E}_b(\vec{r})$ is a solution of $-\vec{\nabla} \times \vec{\nabla}
\times \vec{E}_b(\vec{r}) +\frac{\omega^2}{c^2}\vec{E}_b(\vec{r})=0$
or already assumed to be the space part of a linearly polarized plane wave
$\vec{E}_b(\vec{r})=\vec{E}_0e^{i\vec{k}\vec{r}}$ with a fixed amplitude
vector $\vec{E}_0$. The time dependence $e^{-i\omega t}$ can be omitted in
$\vec{E}_b(\vec{r})$ as well as in $\vec{E}(\vec{r})$.

From (7) a numerical procedure can be deduced if the objects
with $\varepsilon(\vec{r})\ne \varepsilon_0$ only occupy fractions of
space rather small on the scale of the wavelength, not at all principally
necessarily
much smaller than $\lambda$, though. Then we divide them up into finite
elements (Fig.1) of volume $\Delta v$, which we enumerate $i=1,\ldots,N$ and
associate permittivities $\varepsilon(\vec{r}_i)$ or perturbances
$V(\vec{r}_i)=-\frac{\omega^2}{c^2}(\varepsilon(\vec{r}_i)-1)$ and local
fields $\vec{E}(\vec{r}_i)$ uniform over $\Delta v$. The linear sizes of
the object cells should not exceed about $\lambda/10$. They need not at all
be placed on a regular grid. And the only reason for demanding small enough
objects is not to get too many elements $N$.
The integrand in (7) only exists
at places where $V$ does not vanish, and to evaluate the field
$\vec{E}(\vec{r})$ also at such places $\vec{r} \in \{ \vec{r}_i \}$, no
values outside the objects appear in the equation. Changing to finite
elements we thus get a linear system of equations for the fields in the object
cells
\begin{equation}
\vec{E}(\vec{r}_i)=\vec{E}_b(\vec{r}_i)+\sum_{j=1}^N \Delta v\; g(\vec{r}_i,
\vec{r}_j) \; V(\vec{r}_j)\; \vec{E}(\vec{r}_j)
\end{equation}
which can be solved by a matrix inversion. To evaluate the resulting field
at any other place, that is outside the objects, the $\vec{E}(\vec{r}_i)$ just have
to be inserted into the finite-element version of (7):
\begin{equation}
\vec{E}(\vec{r})=\vec{E}_b(\vec{r})+\sum_{j=1}^N \Delta v\; g(\vec{r},
\vec{r}_j) \; V(\vec{r}_j)\; \vec{E}(\vec{r}_j) \qquad 
\vec{r} \not\in \{ \vec{r}_i\}
\end{equation}

Of course, we already needed $g(\vec{r}_i,\vec{r}_j)$ with
$\vec{r}_i=\vec{r}_j$ to set up the system (8). Let us suppose that there
is only a single cell with an $\varepsilon$ differing from the background
$\varepsilon_0$. This is placed into a homogeneous field $\vec{E}_b$.
If the cell has the shape of a sphere, the local field throughout its inside
is aligned in the direction of $\vec{E}_b$ and its value is
$\vec{E}_{loc}=\frac{3}{2+\varepsilon}\vec{E}_b$ \cite{Kop,Jac_3}. No retardation
effects have to be considered here as the size of the cell can in principle
be made arbitrarily small. In (8) only keeping the term of the sum with
$i=j$, setting $\vec{E}(\vec{r}_i)=\vec{E}_{loc}$ and
$V(\vec{r}_i)=-k^2(\varepsilon-1)$ gives
$$\frac{3}{2+\varepsilon}\; \vec{E}_b=\vec{E}_b-\Delta v\;
g(\vec{r}_i,\vec{r}_i) \; k^2(\varepsilon-1) \; \frac{3}{2+\varepsilon}\;
\vec{E}_b$$
from which follows that
\begin{equation}
g(\vec{r}_i,\vec{r}_i)=\frac{1}{3k^2\Delta v}\; {\bf 1}
\end{equation}
The factor 1/3 is also valid for cubic elementary cells, however,
other shapes require different depolarization factors \cite{Kop,Chew}.

The background and the resulting field at each point are already 3-vectors.
Nevertheless, in order to solve (8), imagine the $\vec{E}$s for the object
cells assembled into long or ''double'' vectors of $N$ times 3 components
$$ \vec{\vec{\rm E}} \quad {\rm with} \; (\vec{\vec{\rm E}})_i=\vec{E}(\vec{r}_i)
\qquad {\rm and} \qquad \vec{\vec{\rm E}}_b \quad {\rm with} \quad
(\vec{\vec{\rm E}}_{b})_{i}=\vec{E}_b(\vec{r}_i).$$
With further the big $3N\times 3N$ matrix ${\rm M}$ consisting of 3x3 blocks
$$ M_{ij}={\bf 1}\; \delta_{ij} - \Delta v \; g(\vec{r}_i,\vec{r}_j) \;
V(\vec{r}_j)$$
(8) then reads
\renewcommand{\theequation}{\arabic{equation}a}
\setcounter{equation}{7}  \begin{equation}
{\rm M} \vec{\vec{\rm E}} = \vec{\vec{\rm E}}_b \quad {\rm or} \quad
\vec{\vec{\rm E}}={\rm M}^{-1} \vec{\vec{\rm E}}_b \end{equation}
\renewcommand{\theequation}{\arabic{equation}}
\setcounter{equation}{10}
${\rm M}$ can be inverted using the procedure described in the appendix,
however, as ${\rm M}$ is a full matrix and the complete invers ${\rm M}^{-1}$
is needed, that is of no advantage and a standard inversion algorithm will
do as well.

For the problems of a few small scatterers here by introducing finite elements
the implicit integral equation (7) for the electric field has been turned into
a linear system of equations that is easily solvable. Even shying this effort,
the coarsest, so-called Born approximation consists in replacing $\vec{E}$ in
the integral in (7) or in the sums in (8) or (9) by $\vec{E}_b$. Keeping the
summation over finite elements to estimate the integral we directly get
\begin{equation}
\vec{E}(\vec{r})\approx \vec{E}_b(\vec{r})+\sum_{j=1}^N \Delta v \; g(\vec{r}_i,
\vec{r}_j) \; V(\vec{r}_j) \; \vec{E}_b(\vec{r}_j)
\end{equation}
No distinction between places $\vec{r}$ in or outside the object cells is
necessary in (11). The Born approximation only takes into account first-order
scattering off every object and thus can only be good for weak scatterers
with distances between them rather large on the scale of the wavelength.
Producing a clearly different field pattern from the exact solution including
all scattering orders, Fig.2 demonstrates that the Born approximation is
likely to be insufficient to model near-field optics setups.

\begin{figure}
\includegraphics[width=5.5cm,angle=270]{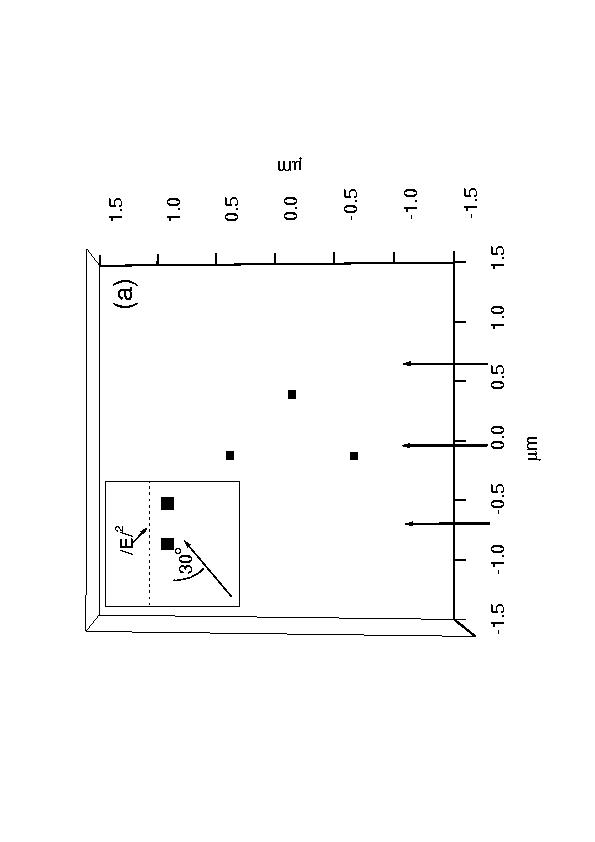}

\includegraphics[width=6.2cm,angle=270]{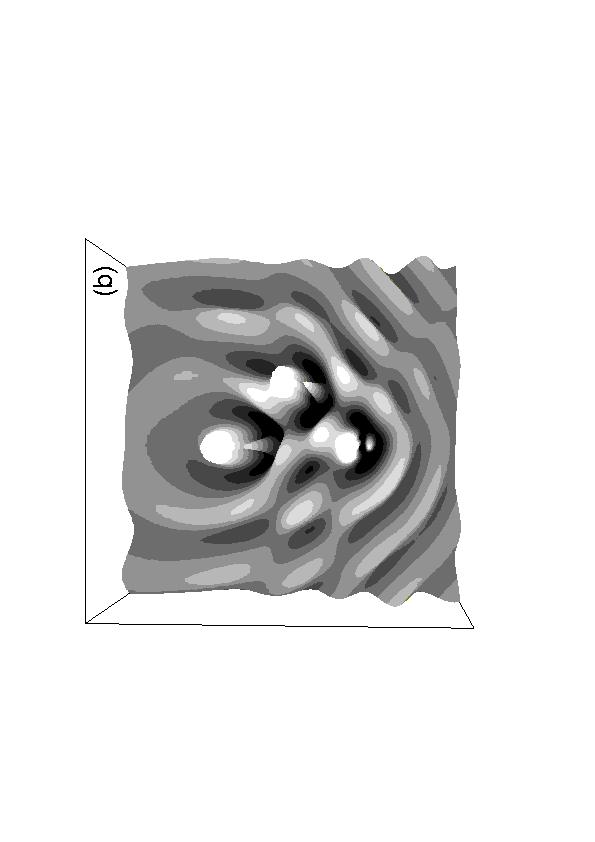}
\includegraphics[width=6.2cm,angle=270]{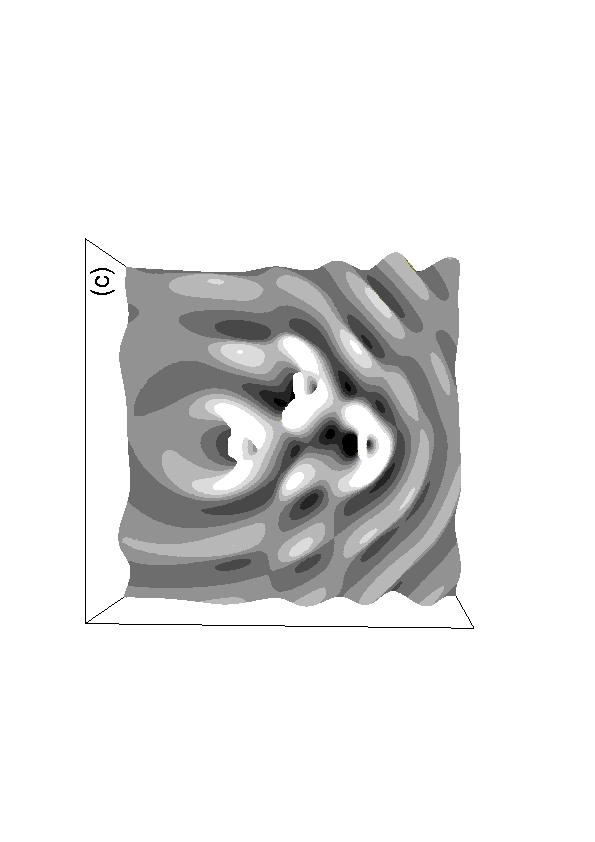}
\caption{Three cubes of 25nm side length and $\varepsilon=-16.0+1.0i$ in
a background of $\varepsilon=1$ illuminated by a p-polarized
plane wave of unit amplitude
with $\lambda=500{\rm nm}$ from the front and under an angle of $30^{\sf o}$
from beneath the plane
defined by the cubes. Setup (a) and electric field intensity $\vert \vec{E}
\vert^2$ calculated via the Born approximation (b) and the exact Green's
functions method (c). Besides subtle differences in the near-field pattern
remark that in (b) the grey-scale is from 0.978 to 1.022 whereas in (c) it is
from 0.995 to 1.005. The intensity map is taken at a height 25nm above the
cubes and given for an area of $3\mu {\rm m} \times 3 \mu {\rm m}$.
}\end{figure}

\subsection{System Green's function and density of states}
Adding an arbitrary source term to our original wave equation (1) changes it into
\begin{equation}
-\vec{\nabla}\times\vec{\nabla}\times \vec{E}(\vec{r})+\frac{\omega^2}{c^2}
\varepsilon(\vec{r})\vec{E}(\vec{r})=\vec{Q}(\vec{r}).
\end{equation}
If now we know a tensor function $G(\vec{r},\vec{r'})$ satisfying
\begin{eqnarray}
-\vec{\nabla}_{\vec{r}}\times\vec{\nabla}_{\vec{r}}\times G(\vec{r},\vec{r'})
+\frac{\omega^2}{c^2}\varepsilon(\vec{r})G(\vec{r},\vec{r'}) = {\bf 1}\;
\delta(\vec{r}-\vec{r'}) \quad {\rm or} \\ 
-\vec{\nabla}_{\vec{r}}\times\vec{\nabla}_{\vec{r}}\times G(\vec{r},\vec{r'})
+k^2 G(\vec{r},\vec{r'}) -V(\vec{r}) G(\vec{r},\vec{r'})
= {\bf 1}\; \delta(\vec{r}-\vec{r'}) \nonumber
\end{eqnarray}
then obviously
\begin{equation}
\vec{E}(\vec{r})=\int d^3\vec{r'} \; G(\vec{r},\vec{r'}) \; \vec{Q}(\vec{r'})
\end{equation}
would give a special solution of (12). Any solution of (1) could be added.
As generally deduced in section I the implicit relation to get $G$ from is the
Dyson equation
\begin{equation}
G(\vec{r},\vec{r'})=g(\vec{r},\vec{r'})+\int d^3\vec{r''} \; g(\vec{r},
\vec{r''}) \; V(\vec{r''}) \; G(\vec{r''},\vec{r'}) 
\end{equation}

$G(\vec{r},\vec{r'})$ for both arguments $\vec{r}$ and $\vec{r'}$ covering all
space is too much information to display at once and usually much more than
what one is interested in.
The imaginary part of $G(\vec{r},\vec{r})$ is proportional to the density of
states $\rho$ \cite{coral,AD}. The deduction of this statement found in quantum mechanics
book \cite{Calla}, however, rather argues with a system of energy eigenstates and the
variation of the Green's function as well as the density of states with
energy. No real $\vec{r}$-space is explicitly mentioned. Our interest lies in
the spatial dependence of the density of states at fixed light frequency
$\omega$.

Even if described in terms of fields, concepts like reactance and work known
from electrical circuits may be applied \cite{Jac_4,circtip}. The time average of the
work done by the fields is given by
\begin{equation}  {\rm Re} \; \frac{1}{2} \int d^3\vec{r'} \;
\vec{J}^*(\vec{r'}) \; \vec{E}(\vec{r'}) \end{equation}
With no other imposed fields, charges or currents than an oscillating point
dipole, the latter will present the only external current $\vec{J}$, which
will thus be located as $\delta(\vec{r'}-\vec{r})$. If the dipole moment
oscillates as $\vec{p}(\vec{r'},t)=\vec{p}_0e^{-i\omega t}
\delta(\vec{r'}-\vec{r})$, the corresponding current is $\vec{J}(\vec{r'},t)
=-i\omega \vec{p}_0e^{-i\omega t} \delta(\vec{r'}-\vec{r})$. Deducing the
wave equation for time harmonic fields (in vacuum for simplicity here) from
Maxwells equations with current term
\begin{equation}
\vec{\nabla}\times \vec{E}-i\omega \mu_0\vec{H}=0 \quad {\rm and} \quad
\vec{\nabla}\times \mu_0 \vec{H}+\varepsilon_0\mu_0 i\omega \vec{E}=\mu_0
\vec{J} \nonumber \end{equation}
leads to \begin{equation}
-\vec{\nabla}\times\vec{\nabla}\times \vec{E}+\frac{\omega^2}{c^2}\vec{E}=
-i\omega\mu_0\vec{J} \end{equation}
from which we see that the source term for the dipole has to be set as
$\vec{Q}(\vec{r'})=-i\omega \mu_0 \vec{J}(\vec{r'})=-\omega^2\mu_0\vec{p}_0
e^{-i\omega t} \delta(\vec{r'}-\vec{r})$. The integral (16) reduces to the
value of $\vec{J}^*\vec{E}$ at $\vec{r}$. The electric field we get from (14):
\begin{equation}
\vec{E}(\vec{r})=-\omega^2\mu_0 e^{-i\omega t} \int d^3\vec{r'} \;
G(\vec{r},\vec{r'}) \; \delta(\vec{r'}-\vec{r}) \; \vec{p}_0 = -\omega^2
\mu_0 e^{-i\omega t} G(\vec{r},\vec{r}) \vec{p}_0
\nonumber\end{equation}
Inserting $\vec{J}^*$ and $\vec{E}$ into (16) the time factors cancel as
expected for a time average and but for a factor $\omega^3\mu_0/2$ we get
\begin{equation}
{\rm Re} \; i \; \vec{p}_0 \; G(\vec{r},\vec{r}) \; \vec{p}_0 =
-\; {\rm Im} \; \vec{p}_0 \; G(\vec{r},\vec{r}) \; \vec{p}_0
\end{equation}
Choosing unit vectors along the coordinate axis for the probe dipole $\vec{p}_0$,
(20) will filter out the trace elements of the matrix $G(\vec{r},\vec{r})$.
We associate $\rho_x(\vec{r})\propto -{\rm Im}G_{xx}(\vec{r},\vec{r})$,
$\rho_y(\vec{r})\propto -{\rm Im}G_{yy}(\vec{r},\vec{r})$,
$\rho_z(\vec{r})\propto -{\rm Im}G_{zz}(\vec{r},\vec{r})$ and a total
$\rho(\vec{r})=\rho_x(\vec{r})+\rho_y(\vec{r})+\rho_z(\vec{r})$.

A motivation for taking the negative imaginary part of $G(\vec{r},\vec{r})$
as a measure for the presence of modes can also be obtained by comparison to
the energy resonance of a forced oscillator \cite{Alonso}. For optimal excitation
from the energy point of view - in contrast to amplitude resonance - the
force has to be $\pi/2$ ahead of the elongation or in phase with the
velocity of the oscillator. $G(\vec{r},\vec{r})$ describes the field caused
by backaction of the system at the place of the probe dipole moment (taken as
reference phase zero), and therefore $-{\rm Im}G(\vec{r},\vec{r})$ is the part
that can in a resonant manner further enhance the dipole oscillation.
(In reality radiation out of the system will provide strong damping.)

We now intend to evaluate a map of $G(\vec{r},\vec{r})$ on, for example, a
horizontal plane. The plane may lie above or below object cells or even cut
some. Like the objects, the, of course, finite area of interest on the plane
is divided into cells. Just depending on the desired resolution of the map the
unit cell length of this mesh may well differ from the cell size chosen to
discretize the objects (Fig.1). The list of object cell midpoints $\{\vec{r}_i\}$
from the last section, which shall be called region $A$, is extended by all
cell midpoints from the map in the plane, which shall be called region $B$
and is now understood to to be included in counting $i$ from 1 to a new $N$.
Analogously to (8) the integral in (15) is replaced by a sum:
\begin{equation}
G(\vec{r}_i,\vec{r}_k)=g(\vec{r}_i,\vec{r}_k)+\sum_{j=1}^N\Delta v\;
g(\vec{r}_i,\vec{r}_j) \; V(\vec{r}_j) \; G(\vec{r}_j,\vec{r}_k)
\end{equation}
It does not matter that the map $B$ has a different mesh from $\Delta v$ as
for the objects $A$, as $V(\vec{r}_j)=0$ for $\vec{r}_j$ in $B$, anyway. (Spatial
overlap of cells from $A$ and $B$ and even coincidence of midpoints is no
problem; a place can be counted with $V(\vec{r}_j)$ in $A$ and without in $B$.)

Although only $G(\vec{r}_i,\vec{r}_i)$ with $\vec{r}_i$ in region $B$ is
wanted as a final result, (21) has to be set up as an equation for a matrix of
all $G$ with each of its arguments any cell in $A$ or $B$, schematically
sketched as
$\begin{pmatrix}\vbox{\halign{\strut # & \vrule \ # \cr
AA & AB \cr \noalign{\hrule} BA & BB \cr }}\end{pmatrix}$.
To solve (21) for $G$ we have to invert the same kind of matrix as in (8), the
only difference being that $j$ now also runs over the plane cells in addition
to the object cells.
\renewcommand{\theequation}{\arabic{equation}a}
\setcounter{equation}{20}
\begin{equation} G={\rm M}^{-1}\; g \end{equation}
\renewcommand{\theequation}{\arabic{equation}}
\setcounter{equation}{21}
$G$ and $g$ are themselves matrices on contrast to vectors $\vec{\vec{E}}$
and $\vec{\vec{E}}_b$. $G$ consists
of 3x3-blocks $G(\vec{r}_j,\vec{r}_k)$, $g$ is made of blocks $g(\vec{r}_i,\vec{r}_k)$
and the 3x3-block at position $(i,j)$ in ${\rm M}$ given by ${\bf 1}\delta_{ij}-
\Delta v\; g(\vec{r}_i,\vec{r}_j) \; V(\vec{r}_j)$. The $g$ in (21a) and ${\rm M}$ like
the big $G$-matrix have the structure
$\begin{pmatrix}\vbox{\halign{\strut # & \vrule \ # \cr
AA & AB \cr \noalign{\hrule} BA & BB \cr }}\end{pmatrix}$.
One could invert ${\rm M}$ as given, for example by the procedure from
appendix A. The matrix $M={\bf 1}-{\rm M}$ there is initialized with $gV$.
Its $AB$ and $BB$ quadrants are zero and will stay zero throughout the procedure,
$M=\begin{pmatrix}\vbox{\halign{\strut # & \vrule \ # \cr
AA & 0 \cr \noalign{\hrule} BA & 0 \cr }}\end{pmatrix}$. This is no contradiction,
as it is not ${\rm M}$ that is singular. Quadrant $BA$ will be needed for
multiplication with $g$ in (21a).
However there is an even more efficient algorithm to get $G$ that
already includes the multiplication by $g$. It directly calculates 
$G=({\bf 1}-gV)^{-1}g$, which is the compact way to write (21a) as the
solution of (15), also denoted $G=g+gVG$ for short. The technical details
can be found in appendix B.

\begin{figure}
\includegraphics[width=6.8cm,angle=270]{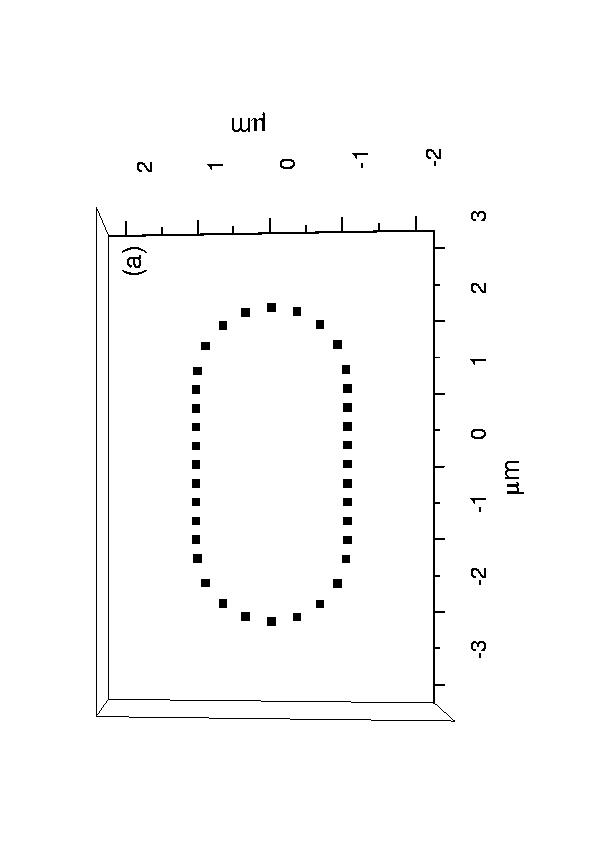}

\vspace{-1.2cm}

\includegraphics[width=7.1cm,angle=270]{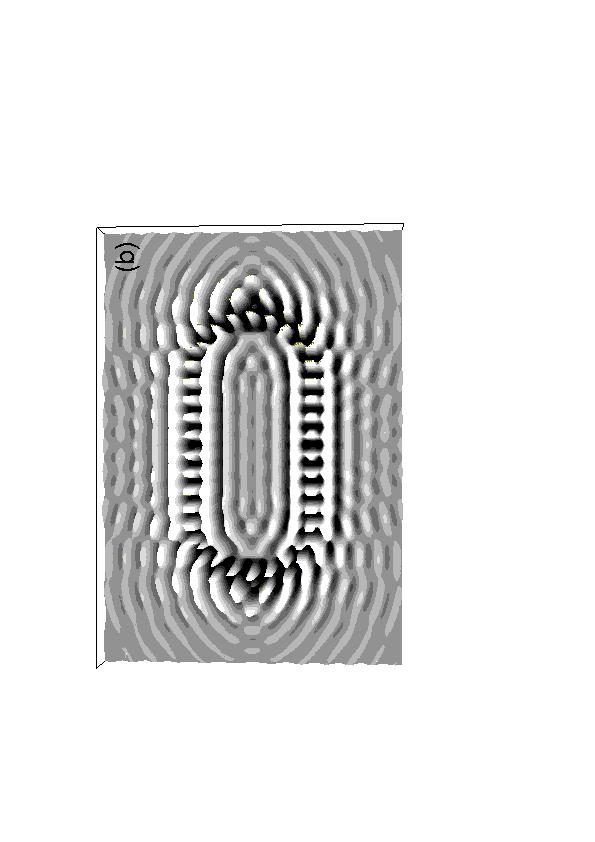}

\vspace{-1.2cm}

\includegraphics[width=7.1cm,angle=270]{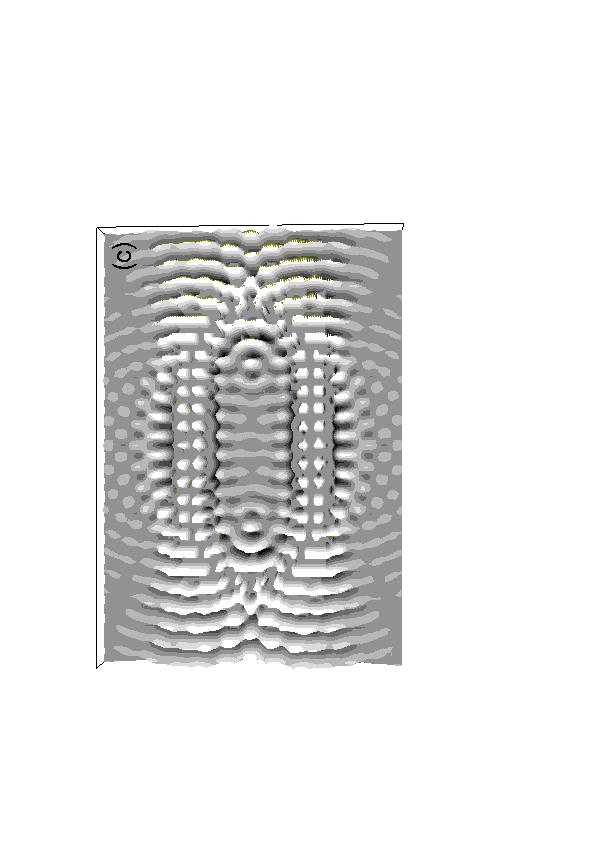}

\vspace{-1.2cm}

\includegraphics[width=7.1cm,angle=270]{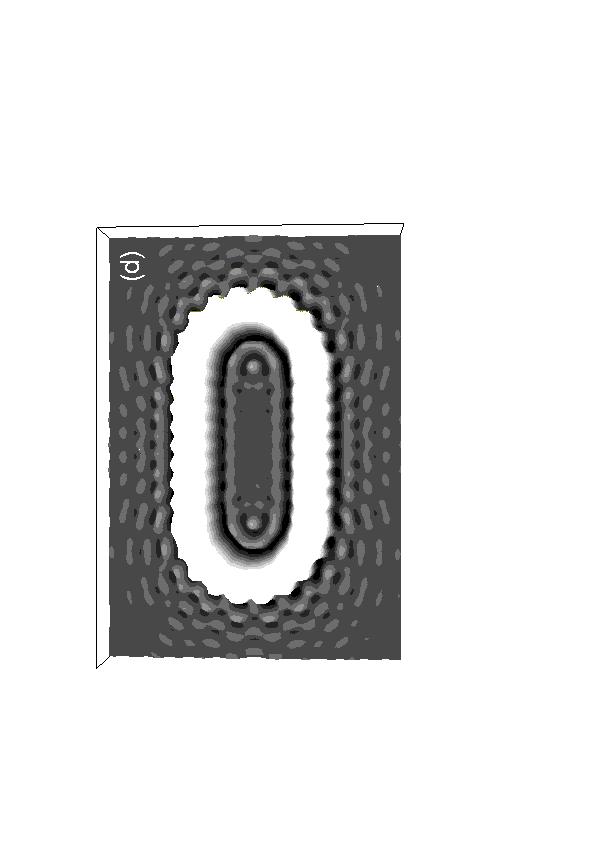}
\end{figure}
\setcounter{figure}{2}
\begin{figure}
\caption{(a) Top view of the structure: gold pads (90x90x30nm)
on a glass subtrate. Calculated densities of states $\rho_{xx}$ (b)
$\rho_{yy}$ (c) and $\rho_{zz}$ (d) in a plane 100nm above the substrate
for wavelength $\lambda=543{\rm nm}$ [reproduced after  \cite{Chicel}].
The grey scale is from -50 to 35 in (b), -45 to 30 in (c), -7.5 to 15 in (d)
and higher or lower values are white or black, respectively.
}\end{figure}

Trace components of $G(\vec{r},\vec{r})$ meaning densities of states for the
three polarization directions (Fig.3)
above an optical coral \cite{coral} in analogy to a quantum coral \cite{Eigler}
have been measured \cite{Chicel} in a so called forbidden-light near-field
optical microscope \cite{forbidden}. The sample consists of a stadium arrangement of gold
particles on a glass surface. The forbidden-light setup prevents detecting
light emitted from the fiber tip that has not passed through surface modes
that make up the density of states for this system.
Like for antinodal and nodal points in a resonator, more energy can go into
the system when the excitation is placed at a point of high density of states
than when coupling is bad where the density is low.

\subsection{Remarks on the source terms and alternative solutions}
In section 4 we saw that it is convenient to start from a solution for the
field in the form (7) if the excitation comes, for example, from a background
field belonging to a plane wave. Though the matrix to invert bore a certain
similarity to the evaluation of the Green's tensor in section 5, with (8) and (9) we
directly calculated the field. In contrast, more adapted to localized sources,
there is (14) as a solution of (12). If there is no additional background field
to cause any excitation, no solution of the equation (1) with zero right
side is to be
added as further contribution and (14) is the field distribution to be
observed. (14) has to be rewritten in terms of finite elements in order to be
used in a numerical calculation. In the same way as the objects the source
$\vec{Q}$ has to be devided into discrete cells or elementary dipoles. To
distinguish their $L$ locations from those of the objects we shall enumerate
them as $\vec{\rho}_{\alpha}$, $\alpha=1,2,\ldots,L$. The place $\vec{r}$ to
evaluate the field $\vec{E}$ may be anywhere outside or inside the objects as
well as beside or even at a source location. For the following development
the Dyson equation for the Green's tensor is needed in a discretized form for
both its variants $G=g+gVG$ and $G=g+GVg$.
\setcounter{equation}{0}
\renewcommand{\theequation}{22\alph{equation}}
\begin{eqnarray}
\vec{E}(\vec{r})&=& \sum_{\alpha=1}^L G(\vec{r},\vec{\rho}_{\alpha})
\vec{Q}(\vec{\rho}_{\alpha}) \\
&=& \sum_{\alpha=1}^L g(\vec{r},\vec{\rho}_{\alpha}) \vec{Q}(\vec{\rho}_{
\alpha}) +\sum_{\alpha=1}^L \sum_{j=1}^N G(\vec{r},\vec{r}_j) V(\vec{r}_j)
g(\vec{r}_j,\vec{\rho}_{\alpha}) \vec{Q}(\vec{\rho}_{\alpha}) \\
&=& \sum_{\alpha=1}^L g(\vec{r},\vec{\rho}_{\alpha}) \vec{Q}(\vec{\rho}_{
\alpha}) + \sum_{\alpha=1}^L \sum_{j=1}^N g(\vec{r},\vec{r}_j) V(\vec{r}_j)
g(\vec{r}_j,\vec{\rho}_{\alpha}) \vec{Q}(\vec{\rho}_{\alpha}) + \nonumber \\
&& \sum_{\alpha=1}^L \sum_{j=1}^N \sum_{i=1}^N g(\vec{r},\vec{r}_i) V(\vec{r}_i)
G(\vec{r}_i,\vec{r}_j) V(\vec{r}_j) g(\vec{r}_j,\vec{\rho}_{\alpha})
\vec{Q}(\vec{\rho}_{\alpha}) \\
&=& \sum_{\alpha=1}^L g(\vec{r},\vec{\rho}_{\alpha})
\vec{Q}(\vec{\rho}_{\alpha}) +\sum_{\alpha=1}^L
\sum_{j=1}^N g(\vec{r},\vec{r}_j) V(\vec{r}_j) G(\vec{r}_j,\vec{\rho}_{\alpha})
\vec{Q}(\vec{\rho}_{\alpha}) \nonumber \\
&=& \sum_{\alpha=1}^L g(\vec{r},\vec{\rho}_{\alpha})\vec{Q}(\vec{\rho}_{\alpha})
+\sum_{j=1}^N g(\vec{r},\vec{r}_j) V(\vec{r}_j) \vec{E}(\vec{r}_j)
\end{eqnarray}
\setcounter{equation}{22}
\renewcommand{\theequation}{\arabic{equation}}
Having in mind a region where and a resolution with which $\vec{E}(\vec{r})$
is to be evaluated like the discretized plane $B$ from the last section, it
would be possible to supply $G$ for all needed combinations of arguments
$(\vec{r},\vec{\rho}_{\alpha})$ and calculate $\vec{E}(\vec{r})$ as the
single sum from (22a). To weave in the influence of the objects, $G$ would have
to be set up as a big matrix like in the last section over all combinations
of three regions $A$, $S$ and $B$ here, the objects, the source and the map.
Having calculated $G$ in the $A$-$B$-scheme from the last section, one
could evaluate
$\vec{E}(\vec{r})$ with $G(\vec{r},\vec{r}_j)$ as written in (22b). However,
the most efficient way is given in (22c). $G(\vec{r}_i,\vec{r}_j)$ is merely
needed for $\vec{r}_i$ and $\vec{r}_j$ from the set of object cells, keeping
a matrix to be inverted as small as possible, namely of $(AA)$-type. Choosing
$\vec{r}_i$ and $\vec{r}_k$ in (21) in the object set $A$, instead of in $A$ or
$B$ as the equation was originally set up for, we see that (21) presents a
closed system of equations for all such $G(\vec{r}_i,\vec{r}_k)$.
Then for (22c) more
summations over products with $g$-functions, which are analytically known for
any pair of arguments, can be considered less demanding in computing time than
the inversion of large matrices.

In the transformation from (22c) to (22d) after swapping index names $i$ and
$j$ in the last sum, $G=g+GVg$ and (14) have been exploited. 
Using (22d) for $\vec{E}(\vec{r})$
renders an implicit equation for the field in the form (10) or (10*) from
section I.
$\vec{E}(\vec{r})$ is the equivalent of $\phi_{ihq}$ and as stated earlier,
we assume that physically there is no background field $\phi_0$ that could
initiate an additional field distribution $\phi_{ih}$.
Although not very convenient, a plane wave as exciting field could be
understood as stemming from a sufficiently long and dense array of Huygens
elementary dipole sources $\vec{Q}$ reasonably far away from the objects.
The other way round,
for a single dipole source or a number of dipole sources distributed in space
the field they would produce at any location $\vec{r}$ in homogeneous space
is the superposition of their individual fields, and putting
$\vec{E}_b(\vec{r})=\sum_{\alpha=1}^L g(\vec{r},\vec{\rho}_{\alpha})
\vec{Q}(\vec{\rho}_{\alpha})$ the ansatz (7) can be used also for this case.

Like the field $\vec{E}$ anywhere was obtained as a straight-forward summation
once having its values at the places of the object cells, finally an alternative
way to the procedure from the last section to get the Green's tensor $G$ shall
be given, also requiring only the inversion of a matrix with size the number
of object cells. Series expansion is used to rewrite the solution of the
Dyson equation:
\setcounter{equation}{0}
\renewcommand{\theequation}{23\alph{equation}}
\begin{eqnarray}
G&=& ({\bf 1}-gV)^{-1}g = [{\bf 1}+\sum_{n=1}^{\infty} (gV)^n]g 
= g+gV(\sum_{n=0}^{\infty} (gV)^n)g \nonumber \\
&=& g+gV({\bf 1}-gV)^{-1}g \\
&=& g+gVg+gV(\sum_{n=0}^{\infty} (gV)^n)gVg =
g+gVg+gV\; ({\bf 1}-gV)^{-1}g\; Vg \nonumber  \\
&=& g+gVg+gVGVg
\end{eqnarray}
Designating regions the spatial arguments belong to on (23a) we get for
$G_{BB}$: \setcounter{equation}{0}
\renewcommand{\theequation}{23\alph{equation}'}
\begin{equation}
G_{BB}=g_{BB}+g_{BA}\; V_A\; ({\bf 1}-gV)^{-1}_{AA}\; g_{AB},
\end{equation}
As $V$ does not vanish only in region $A$, the first index of
$({\bf 1}-gV)^{-1}$ obviously must be $A$. $({\bf 1}-gV)^{-1}$ being
$\sum_{n=0}^{\infty} (gV)^n$, for power zero the second region index automatically
is the same as the first and all other powers ending with $V$ imply second
index $A$.
It is sufficient to set up the matrix ${\bf 1}-gV$ as an $(AA)$-block
and invert that. Should one prefer to evaluate a complete $G$, which
differs from the above inverted matrix by a factor $g$, line
(23b) like (22c) shows that it is in principle only necessary to get $G$ from
some self-consistent implicit equation in the object region $A$.
$G_{AA}$ can be constructed applying the procedure described in appendix B to
a matrix set up as $(AA)$-block only. The inversion has to be completed in this
case, though. Going through the diagonal elements, all lines and columns
have to be updated in each step, including the ones above and to the left of as well as
the ones the respective diagonal element is in. Writing (23b)
\setcounter{equation}{1}
\renewcommand{\theequation}{23\alph{equation}'}
\begin{equation}
G_{BB}=g_{BB}+g_{BA}\; V_A\; g_{AB}+g_{BA}\, V_A\, G_{AA}\; V_A\; g_{AB}
\end{equation}
as summation over discrete elements ready for use in a calculation then reads:
\setcounter{equation}{1}
\renewcommand{\theequation}{23\alph{equation}''}
\begin{eqnarray}
G(\vec{r},\vec{r'})=g(\vec{r},\vec{r'})&+&\sum_{j=1}^N g(\vec{r},\vec{r}_j)
\; V(\vec{r}_j) \; g(\vec{r}_j,\vec{r'})  \nonumber \\
&+& \sum_{j=1}^N \sum_{i=1}^N g(\vec{r},\vec{r}_j) \; V(\vec{r}_j) \;
G(\vec{r}_j,\vec{r}_i) \; V(\vec{r}_i) \; g(\vec{r}_i,\vec{r'})
\end{eqnarray}
Of course, summations run over all object cells here. No numerical advantage
can be drawn out of $\vec{r'}=\vec{r}$ in (23b''). With the same effort of making
$G_{AA}$ it can be used to evaluate maps of $G(\vec{r},\vec{r})$ as well as
plots of $G(\vec{r},\vec{r'})$ with $\vec{r'}$ fixed or even some function of
$\vec{r}$.
\setcounter{equation}{23}
\renewcommand{\theequation}{\arabic{equation}}

\subsection{Conclusions and Outlook for section II}
We have presented a method to solve the problem of scattering of electromagnetic
waves off an arbitrary distribution of dielectric objects, that is the exact
evaluation of the field, especially in the near zone where higher-order
multiple reflections can become important. Besides the field distribution
we have obtained the Green's tensor characterizing the system independently
from the form of the excitation. It represents the response function and
also the density of states for supported electric fields.

For a methodical introduction we have restricted our considerations to
dielectric materials and the electric field. Without magnetic susceptibilities
the magnetic field distribution can be calculated once having the electric
field inside the objets by a formula like (9) with the magnetic background
field and replacing $g$ by a tensor including the conversion from the electric
to the magnetic field by taking the rotation \cite{Girard}. It is further possible to treat
non-uniform magnetic permeabilities and even mixed systems with dielectric
and magnetic objects \cite{EPJel}. The electric Green's tensor presented above is then
paired by a magnetic counterpart and genuine mixed response functions also
exist. Whereas the calculation of the field distributions even for mixed
systems is quite straight forward, the construction of the Green's tensor
is more involved. It lives of the idea of handling one kind of objects first
and then considering this setup as the background to include the other kind.
There is no approximation or ranking in importance in this procedure.

The discussion here has only considered finite objects in a homogeneous
background as well as cartesian coordinates where vector and tensor components
have been written out. Cylindrical and spherical coordinates are also commonly
used \cite{Chew,ellipse}
and the Green's functins formalism has been developped for layered media
\cite{MarLith,Chew,stratified}.
Besides wave-guide applications the use for modelling typical near-field
optics experiments, where the microstructures to investigate are prepared on
a substrate surface, lies in putting the influence of this surface into a
background Green's tensor \cite{Girard,AD} which is then implied the way we used $g$ here.

Details of applications of the Green's functions technique in electrodynamics
to more complicated situations as well as beautiful results of corresponding
experiments can be found in the given references. This text focussed on
calculation techniques and further intended to give an overview of slightly
different formal ways to calculate Green's tensors and fields of which either
may be optimal for a specific problem.

\subsection*{Appendix A: an unusual matrix inversion}
Suppose a complex quadratical matrix to invert is already given in the form
${\rm M} ={\bf 1}-M={\bf 1}-(m)_{ij}$ or if it is not, we rewrite it like that.
There is no restriction on the values of the numbers $m_{ij}$. To get the
inverted matrix proceed as follows:
Of matrix $M$ one by one take the diagonal elements $m_{ii}$ and to all elements
$m_{ab}$ add $m_{ai}(1-m_{ii})^{-1}m_{ib}$. After having worked through the
matrix for one such $m_{ii}$, the changed matrix values have to be taken to do
so for the next, also already changed, diagonal element. Obviously $N$ such
steps are required for an $N\times N$-matrix. This will yield $({\bf 1}-M)^{-1
}-{\bf 1}$, such that in the end 1 has to be added to all diagonal elements in
order to obtain $({\bf 1}-M)^{-1}$. For clearness we write out the first two
transformation steps of the matrix:
\begin{eqnarray}
\begin{pmatrix} m_{11} & m_{22} & \cr m_{21} & m_{22} & \cr & & \ddots \cr 
\end{pmatrix} \rightarrow \begin{pmatrix}
m_{11}+m_{11}(1-m_{11})^{-1}m_{11} & m_{12}+m_{11}(1-m_{11})^{-1}m_{12} \cr
m_{21}+m_{21}(1-m_{11})^{-1}m_{11} & m_{22}+m_{21}(1-m_{11})^{-1}m_{12} \cr 
\end{pmatrix} \nonumber \\
\rightarrow \begin{pmatrix}
m_{11}+m_{11}(1-m_{11})^{-1}m_{11} + & m_{12}+m_{11}(1-m_{11})^{-1}m_{12} +\cr
(m_{12}+m_{11}(1-m_{11})^{-1}m_{12})\cdot & (m_{12}+m_{11}(1-m_{11})^{-1}m_{12})\cdot \cr
(1-m_{22}-m_{21}(1-m_{11})^{-1}m_{12})^{-1} & (1-m_{22}-m_{21}(1-m_{11})^{-1}m_{12})^{-1} \cr
\cdot (m_{21}+m_{21}(1-m_{11})^{-1}m_{11}) &  \cdot (m_{22}+m_{21}(1-m_{11})^{-1}m_{12}) \cr
& \cr
m_{21}+m_{21}(1-m_{11})^{-1}m_{11} + & m_{22}+m_{21}(1-m_{11})^{-1}m_{12} +\cr
(m_{22}+m_{21}(1-m_{11})^{-1}m_{12})\cdot & (m_{22}+m_{21}(1-m_{11})^{-1}m_{12})\cdot \cr
(1-m_{22}-m_{21}(1-m_{11})^{-1}m_{12})^{-1} & (1-m_{22}-m_{21}(1-m_{11})^{-1}m_{12})^{-1} \cr
\cdot (m_{21}+m_{21}(1-m_{11})^{-1}m_{11}) &  \cdot (m_{22}+m_{21}(1-m_{11})^{-1}m_{12}) \cr
\end{pmatrix} \rightarrow \ldots \nonumber \\
\end{eqnarray}
The inverted matrix $({\bf 1}-M)^{-1}$ can be represented as a geometric series:
\begin{equation}
({\bf 1}-M)^{-1}={\bf 1}+M+M^2+\ldots
\end{equation}
Truncating and using the sum from the right side is only possible if the series
converges whereas the closed form on the left is valid in any case.
In contrast to the infinite sum on the right side of (25), our inversion procedure
consists in a finite number of steps of adding contributions to the matrix
elements. Nevertheless, (25) tells us that the invers $({\bf 1}-M)^{-1}$ is the
sum of all powers of $M$ and thus each element in row $a$ and column $b$ must be
the sum of all possible products $m_{an_1}m_{n_1n_2}m_{n_2n_3}\ldots m_{n_{j-1
}n_j}m_{n_jb}$ with any number $j$ of inner indices including none. (Diagonal
elements get an extra +1.) There are $N$ different indices $n_i$ and they may
repeat, of course. Considering that $(1-m_{ii})^{-1}$ can also be written as
$$
(1-m_{ii})^{-1}=1+m_{ii}+m_{ii}^2+\ldots =1+m_{ii}+m_{ii}m_{ii}+\ldots
$$
we see that the first step in (24) adds to each matrix element $m_{ab}$ the sum
of all products $m_{a1}m_{11}m_{11}\ldots m_{11}m_{1b}$. In these at
least one pair of indices 1 is squeezed between $a$ and $b$ as in $m_{a1}m_{1b}$,
the contribution $m_{ab}$ was already there. In the second step all products
$m_{a\nu_1}m_{\nu_1\nu_2} \ldots m_{\nu_{\alpha}2}m_{2\nu_{\beta}} \ldots
m_{\nu_{j-1}\nu_j}m_{\nu_jb}$ with
$\nu_1\nu_2\ldots \nu_{j-1}\nu_j$ every possible sequence of 1s and 2s will be
added. The products with only indices 1 between $a$ and $b$ were there before.
In the third step every sequence of indices 1, 2 and 3 with at least one
3-link is added. And so on until in the end at each
matrix position between outer indices $a$ and $b$ we have created all possible
sequences of an endless game of dominos $m_{ij}$ with numbers $i$ and $j$ from
1 to $N$. This argument was to proove that the result of (24) indeed gives
$({\bf 1}-M)^{-1}-{\bf 1}$. We calculate a finite number of $(1-m_{ii})^{-1}$
or products $m_{ai}(1-m_{ii})^{-1}m_{ib}$. The sequence on the right side of (25)
need not converge and the original entries in $M$ need not at all be small
compared to 1 in their absolute values. There is no approximation in the sense
of a perturbation theory. If the matrix ${\bf 1}-M$ is degenerate, the failure
of the inversion will be noticed when a value $1-m_{ii}$ becomes zero at some
step. Not to confuse notation, remark that in (24) and in
products in the text like $m_{a1}m_{11}\ldots m_{1b}$ letters $m$ meant the
original matrix entries whereas in expressions $1-m_{ii}$, $(1-m_{ii})^{-1}$ and
$m_{ai}(1-m_{ii})^{-1}m_{ib}$ we referred to the entries at the respective step
of the matrix transformation.

In the application from the main text $i=1,\ldots ,N$ enumerates the object cells.
At position $(i,j)$ in $({\bf 1}-gV)^{-1}$ when expanded into a series having
every possible sequence
$g(\vec{r}_{n_1},\vec{r}_{n_2})V(\vec{r}_{n_2})g(\vec{r}_{n_2},\vec{r}_{n_3})
V(\vec{r}_{n_3})g \ldots g(\vec{r}_{n_{j-1}},\vec{r}_{n_j})V(\vec{r}_{n_j})$
shows that the resulting field at any place (inside or outside the objects)
is the interference of the background field and the fields reradiated by all
the object dipoles having undergone every possible scattering path between
the objects (Fig.1).
A complication in the electrodynamics application at this stage is the fact that each
matrix element $m$ actually in itself is a 3x3 matrix indicating the effect
of three field components at one place onto three field components at a another place.
$g$ on the discrete space of the objects can be written blockwise with a
scheme of quadruple indices

\begin{center}
\begin{picture}(17.2,10)
\put(0,6.5){\dashbox{0.2}(7.75,4){ $\begin{matrix}
\{ \vec{r}_1,x;\vec{r}_1,x\} &
\{ \vec{r}_1,x;\vec{r}_1,y\} &
\{ \vec{r}_1,x;\vec{r}_1,z\} \cr &&\cr
\{ \vec{r}_1,y;\vec{r}_1,x\} &
\{ \vec{r}_1,y;\vec{r}_1,y\} &
\{ \vec{r}_1,y;\vec{r}_1,z\} \cr &&\cr
\{ \vec{r}_1,z;\vec{r}_1,x\} &
\{ \vec{r}_1,z;\vec{r}_1,y\} &
\{ \vec{r}_1,z;\vec{r}_1,z\} \cr \end{matrix}$ }}

\put(8,6.5){\makebox(7.75,4){ $\begin{matrix}
\{ \vec{r}_1,x;\vec{r}_2,x\} &
\{ \vec{r}_1,x;\vec{r}_2,y\} &
\{ \vec{r}_1,x;\vec{r}_2,z\} \cr &&\cr
\{ \vec{r}_1,y;\vec{r}_2,x\} &
\{ \vec{r}_1,y;\vec{r}_2,y\} &
\{ \vec{r}_1,y;\vec{r}_2,z\} \cr &&\cr
\{ \vec{r}_1,z;\vec{r}_2,x\} &
\{ \vec{r}_1,z;\vec{r}_2,y\} &
\{ \vec{r}_1,z;\vec{r}_2,z\} \cr \end{matrix}$ }}

\put(0,1.75){\makebox(7.75,4){ $\begin{matrix}
\{ \vec{r}_2,x;\vec{r}_1,x\} &
\{ \vec{r}_2,x;\vec{r}_1,y\} &
\{ \vec{r}_2,x;\vec{r}_1,z\} \cr &&\cr
\{ \vec{r}_2,y;\vec{r}_1,x\} &
\{ \vec{r}_2,y;\vec{r}_1,y\} &
\{ \vec{r}_2,y;\vec{r}_1,z\} \cr &&\cr
\{ \vec{r}_2,z;\vec{r}_1,x\} &
\{ \vec{r}_2,z;\vec{r}_1,y\} &
\{ \vec{r}_2,z;\vec{r}_1,z\} \cr \end{matrix}$ }}

\put(8,1.75){\dashbox{0.2}(7.75,4){ $\begin{matrix}
\{ \vec{r}_2,x;\vec{r}_2,x\} &
\{ \vec{r}_2,x;\vec{r}_2,y\} &
\{ \vec{r}_2,x;\vec{r}_2,z\} \cr &&\cr
\{ \vec{r}_2,y;\vec{r}_2,x\} &
\{ \vec{r}_2,y;\vec{r}_2,y\} &
\{ \vec{r}_2,y;\vec{r}_2,z\} \cr &&\cr
\{ \vec{r}_2,z;\vec{r}_2,x\} &
\{ \vec{r}_2,z;\vec{r}_2,y\} &
\{ \vec{r}_2,z;\vec{r}_2,z\} \cr \end{matrix}$ }}

\put(16,0.5){\makebox(1,1){$\ddots $}}

\put(0,1.5){\line(1,0){17.2}}
\put(0,6.2){\line(1,0){17.2}}
\put(7.85,0){\line(0,1){10}}
\put(16,0){\line(0,1){10}}

\put(1.45,10){\oval(2.5,0.68)}
\put(3.9,8.5){\oval(2.5,0.68)}
\put(6.35,7){\oval(2.5,0.68)}

\put(9.45,5.15){\oval(2.5,0.68)}
\put(11.9,3.65){\oval(2.5,0.68)}
\put(14.35,2.15){\oval(2.5,0.68)}

\end{picture}
\end{center}

$V(\vec{r}_k)$ in the product $g(\vec{r}_i,\vec{r}_k)V(\vec{r}_k)$
just multiplies the respective column.

One could use the inversion procedure working off the diagonal elements
marked by ovals, requiring $3N$ steps then. However, the process is equally
applicable to 3x3 blocks as marked by the dashed rectangles,
since by its dimension the whole matrix can be
divided up into 3x3-blocks. Then $({\bf 1}-m_{ii})^{-1}$ means the inversion of
a 3x3 matrix and $m_{ai}({\bf 1}-m_{ii})^{-1}m_{ib}$ to update the blocks
means the product of three 3x3 matrices. These operations should be
programmed as elementary procedures. 

The given procedure to invert a matrix can become of advantage if for sparse
matrices conventional routines run into numerical difficulties because of
many zero values. Besides that, it can be adapted to become quite efficient
if only parts of the inverted matrix are needed or for symmetry reasons it is
known that blocks or patterns of matrix elements vanish and will stay zero
throughout the inversion. Although for the calculation of the Green's tensor
a slightly modified procedure is applied that directly optimizes the numerical
solution of the Dyson equation (see appendix B), the matrix inversion was
discussed here, because it may be used for more general purposes and in other
contexts as well.

\subsection*{Appendix B: Calculating the Green's tensor}
In the following instructions are given how to calculate the Green's tensor
\setcounter{equation}{20}
\begin{equation}
G=({\bf 1}-gV)^{-1}g=g+gVg+gVgVg+\ldots
\end{equation}
\setcounter{equation}{25}
being efficient in the way that finally only $G(\vec{r},\vec{r})$ of equal
arguments on the mesh points of map $B$ need to have the correct values
\cite{priv,EPJel}. For a start nevertheless consider that $G(\vec{r}_i,\vec{r}_k)$
with arguments $\vec{r}_i$ and $\vec{r}_k$ from the big set of all object
and map cell midpoints will have to be the sum of all products
$$
g(\vec{r}_i,\vec{r}_{n_1})V(\vec{r}_{n_1})g(\vec{r}_{n_1},\vec{r}_{n_2})
V(\vec{r}_{n_2})g(\vec{r}_{n_2},\vec{r}_{n_3}) \ldots V(\vec{r}_{n_{j-1}},
\vec{r}_{n_j})g(\vec{r}_{n_j},\vec{r}_k).
$$
$\vec{r}_{n_1}$, $\vec{r}_{n_2}$, $\ldots$, $\vec{r}_{n_j}$ can only be object cells and
any sequence of them has to be created, including the empty one with no $V$
giving the term $g(\vec{r}_i,\vec{r}_k)$. Initiate a matrix
$\begin{pmatrix}
\vbox{\halign{\strut # & \vrule \ # \cr
AA & AB \cr \noalign{\hrule} BA & {\tiny
\vbox{\halign{\strut # & # & # \cr
BB & & \cr & $\ddots$ & \cr & & BB \cr }} }
\cr }} \end{pmatrix}$
- for simplicity call it $G$ from the beginning - with line and column
arguments running over all object cells in $A$ and all map cells in $B$ with
$g(\vec{r}_i,\vec{r}_k)$ in each 3x3-subblock. In the $BB$-quadrant only
diagonal blocks $g(\vec{r}_i,\vec{r}_i)$ will be needed, however.
Work off the diagonal subblocks through the $AA$-quadrant.
For the $n$th one prepare the inverted 3x3-matrix $({\bf 1}-G(\vec{r}_n,
\vec{r}_n)V(\vec{r}_n))^{-1}$ and to all entries from line $n+1$ and column
$n+1$ on add the product given below.
\begin{equation}
G(\vec{r}_i,\vec{r}_k) \rightarrow G(\vec{r}_i,\vec{r}_k) +
G(\vec{r}_i,\vec{r}_n) \; V(\vec{r}_n)
\; ({\bf 1}-G(\vec{r}_n,\vec{r}_n)V(\vec{r}_n))^{-1}
\; G(\vec{r}_n,\vec{r}_k)
\end{equation}
In the following step use the updated entries in the above recipe. In the
$n$th step you need not update entries in line $n$ or above or in column $n$
or to the left of it, because these will not be needed as multiplication
factors $G(\vec{r}_i,\vec{r}_m)$ and $G(\vec{r}_m,\vec{r}_k)$ for $m>n$
any more. The parts of quadrants $AB$ and $BA$ remaining after these
restrictions have to be changed by (26) for every $n$.
Merely diagonal 3x3 blocks have to be done in the $BB$-quadrant.
The first step adds all products consisting of any non-zero number of
factors $V(\vec{r}_1)$ between $g$s, the second step all products eventually
containing $V(\vec{r}_1)$ and one up to any number of $V(\vec{r}_2)$, and
so on. The procedure is finished after $\vec{r}_n$ has run down the
diagonal of the $AA$-quadrant. All sequences of multiple scattering from
the objects are then included.

We could have filled
the whole $BB$-quadrant with initial $g_{BB}$ 3x3-blocks and updated the
entire $BB$-quadrant in each step. Like in the matrix inversion procedure
from appendix A we should further have worked down the complete diagonal and
done (26) for every $\vec{r}_n$ from region $B$ as well. There will, however,
be no additions as $V(\vec{r}_n)=0$ for $\vec{r}_n$ in $B$ (even if a map
cell accidently coincides with an object cell). This argument also reveals
why non-diagonal elements in $BB$ do not have to be evaluated. They can
never appear as multiplication factors $G(\vec{r}_i,\vec{r}_n)$ or
$G(\vec{r}_n,\vec{r}_k)$. The $BB$ diagonal is the resulting $G(\vec{r},
\vec{r})$-map we wanted.

Initiating the matrix by $g$ and introducing the $V$ on treating the
respective diagonal element makes the outcome of the procedure directly
$({\bf 1}-gV)^{-1}g$ compared to initiating the matrix with $gV$ and as an
intermediate step obtaining $({\bf 1}-gV)^{-1}$.

\newpage
\setcounter{equation}{0}

\section{All-order quantum transport}

It is demonstrated how the transport problem for two open free-electron gas
reservoirs with arbitrary coupling can be solved by finding the system's
Green's function. In this sense the article is an introduction on Green's
functions for treating interaction.
A very detailed discussion of the current formula is given on an elementary
basis.
Despite formal resemblances the
stationary transport situation, however, differs in its nature from
introducing coupling between energy levels in a closed system where then
the interest lies in modified eigenvalues and eigenstates.

\subsection{Introduction}
By modern lithography techniques so-called point contacts \cite{Agrait} can be arranged
between conductors. These have sufficiently small dimensions such that
electronic modes get quantized. However, coupling across such constrictions
need not be so weak as to be described by a small tunnel probability, but
can be influenced by coherent interference of multiple reflections. Point
contacts can be obtained by indenting STM-tips into some material \cite{RubioSTM},
electromigration \cite{herre} or the break-junction technique \cite{break}. Whereas for constrictions
imposed by gate electrodes to a two-dimensional electron gas in semiconductors
\cite{vanWees}
one observes quantized conductance values in the sense of fully transmitting
or totally switched-off modes, the application in mind behind this work is
the type of connection like the single-atom contact, characterized by an
ensemble of channels \cite{buett}, which can also have intermediate transmission amplitudes
between zero and one \cite{Sche97}. Even with some fully transmitting modes the contact
bears a resistance in the order of the quantum resistances $R_k=h/e^2=26k\Omega$
\cite{SCT}, such
that viewing the system as a left and a right side with some interaction is
an appropriate picture. Furthermore contributions from several channels just
add in the current. Although a transmitting channel in a point contact is the
application in mind, this article shows how to set up a general procedure
to solve the problem of transport for two open reservoirs with more or less
strong coupling between them, and thus how the Green's functions formalism
from section I is implied in an area of current research
\cite{aktuel_1,aktuel_2}.

\subsection{Green's functions formalism}
As a preparation, consider systems like, for example, the bulk material on the
left or the right side (Fig.1a) without coupling. For these we suppose that we
know the Hamiltonian $H^0$ and the wave function $\Psi^0$ at any given energy
$\hbar\omega$ satisfying the Schr\"odinger equation. Denoting the Hamiltonians
$H_{LL}=H_{RR}=H^0$ (doubling the index makes sense later), the wave functions
$\Psi_L^0=\Psi_R^0=\Psi^0$ and the time derivative as $\partial_{\tau}$, the
Schr\"odinger equations for both sides - for the moment just formally put
into matrix form - are
\begin{equation}
\begin{pmatrix} i\hbar\partial_{\tau}-H_{LL} & 0 \cr 0 & i\hbar\partial_{\tau}-H_{RR} \cr\end{pmatrix}
\begin{pmatrix}\Psi^0_L(\tau) \cr \Psi^0_R(\tau)\cr\end{pmatrix}=\begin{pmatrix}0 \cr 0\cr \end{pmatrix}
\end{equation}
Corresponding to this differential equation we have the Green's functions
equation
\begin{eqnarray}
\begin{pmatrix} i\hbar\partial_{\tau}-H_{LL} & 0 \cr 0 & i\hbar\partial_{\tau}-H_{RR} \cr\end{pmatrix}
\begin{pmatrix} g_{LL}(\tau,\tau') & 0 \cr 0 & g_{RR}(\tau,\tau')\cr \end{pmatrix}=
\begin{pmatrix} \delta(\tau-\tau') & 0 \cr 0 & \delta(\tau-\tau')\cr \end{pmatrix}
\nonumber \\ \end{eqnarray}
If we knew $g=g_{LL}=g_{RR}$, we could immediately also give a solution to (1)
with a source term added
\begin{equation}
\begin{pmatrix} i\hbar\partial_{\tau}-H_{LL} & 0 \cr 0 & i\hbar\partial_{\tau}-H_{RR} \cr \end{pmatrix}
\begin{pmatrix} \Psi^q_L(\tau) \cr \Psi^q_R(\tau)\cr \end{pmatrix}=
\begin{pmatrix} Q_L(\tau) \cr Q_R(\tau)\cr \end{pmatrix}
\end{equation}
namely
\begin{equation}
\begin{pmatrix} \Psi^q_L(\tau) \cr \Psi^q_R(\tau)\cr \end{pmatrix}=
\int d\tau' \; \begin{pmatrix} g_{LL}(\tau,\tau') & 0 \cr 0 & g_{RR}(\tau,\tau')\end{pmatrix} \;
\begin{pmatrix} Q_L(\tau') \cr Q_R(\tau')\cr \end{pmatrix}
\end{equation}
However, we are more interested in the solution when an interaction between
left and right is present, expressed through coupling Hamiltonians $H_{LR}$
and $H_{RL}$,
\begin{equation}
\begin{pmatrix} i\hbar\partial_{\tau}-H_{LL} & -H_{LR} \cr -H_{RL} & i\hbar\partial_{\tau}-H_{RR} \cr\end{pmatrix}
\begin{pmatrix} \Psi_L(\tau) \cr \Psi_R(\tau)\cr\end{pmatrix}=\begin{pmatrix} 0 \cr 0\cr\end{pmatrix}
\end{equation}
and we shall refer to this case by $\Psi$ without upper index. Introducing
the coupling, of course, was the motivation for writing the Schr\"odinger
equation in matrix form over the site space consisting of
L(left) and R(right). Putting the coupling terms on the
right side of the equation, they mimic a source
\begin{equation}
\begin{pmatrix} i\hbar\partial_{\tau}-H_{LL} & 0 \cr 0 & i\hbar\partial_{\tau}-H_{RR} \cr\end{pmatrix}
\begin{pmatrix} \Psi_L(\tau) \cr \Psi_R(\tau)\cr\end{pmatrix}=
\begin{pmatrix} 0 & H_{LR} \cr H_{RL} & 0 \cr\end{pmatrix}
\begin{pmatrix} \Psi_L(\tau) \cr \Psi_R(\tau)\cr\end{pmatrix}
\end{equation}
and following (4) the solution can formally be written as
\begin{eqnarray}
\begin{pmatrix} \Psi_L(\tau) \cr \Psi_R(\tau)\cr\end{pmatrix}=\int d\tau'\;
\begin{pmatrix} g_{LL}(\tau,\tau') & 0 \cr 0 & g_{RR}(\tau,\tau')\end{pmatrix}
\begin{pmatrix} 0 & H_{LR} \cr H_{RL} & 0 \cr\end{pmatrix}
\begin{pmatrix} \Psi_L(\tau') \cr \Psi_R(\tau')\cr\end{pmatrix} \nonumber \\
\end{eqnarray}
obtaining the implicit Lippmann-Schwinger equation for the wave functions.
Although for the coupled system we do not expect the eigenvectors of
$\begin{pmatrix} H_{LL} & H_{LR} \cr H_{RL} & H_{RR} \cr\end{pmatrix}$ to be one with only an
upper component localized on the left and one with only a lower component
localized on the right, it is convenient to denote these solutions as vectors
$\begin{pmatrix} \Psi_L \cr \Psi_R \cr \end{pmatrix}$, keeping indices
L and R. In contrast
to section II, where the Lippmann-Schwinger equation was solved in a discretized
form to obtain field distributions, for our purposes here (7) will merely
serve as a formal step in the derivation of the Green's function. What we are
precisely interested in from a physical point of view is the current that will
flow between the left and the right side, and not necessarily explicitly
evaluating the wave functions.

The Green's function $G$ of the coupled system is inferred from an equation
analogous to (5), namely
\begin{eqnarray}
\begin{pmatrix} i\hbar\partial_{\tau}-H_{LL} & -H_{LR} \cr -H_{RL} & i\hbar
\partial_{\tau}-H_{RR} \cr\end{pmatrix} \begin{pmatrix} G_{LL}(\tau,\tau') & G_{LR}(\tau,\tau') \cr
G_{RL}(\tau,\tau') & G_{RR}(\tau,\tau') \cr\end{pmatrix} = \begin{pmatrix} \delta(\tau-\tau') &
0 \cr 0 & \delta(\tau-\tau')\cr\end{pmatrix} \nonumber \\
\end{eqnarray}
$G$ like the Hamiltonian is a full matrix in site space. With $G$, a solution of
\begin{equation}
\begin{pmatrix} i\hbar\partial_{\tau}-H_{LL} & -H_{LR} \cr -H_{RL} & i\hbar
\partial_{\tau}-H_{RR} \cr \end{pmatrix} \begin{pmatrix} \Psi^{cq}_L(\tau) \cr \Psi^{cq}_R(\tau)\cr\end{pmatrix} =
\begin{pmatrix} Q_L(\tau) \cr Q_R(\tau) \cr\end{pmatrix} 
\end{equation}
would read
\begin{equation}
\begin{pmatrix} \Psi^{cq}_L(\tau) \cr \Psi^{cq}_R(\tau)\cr\end{pmatrix} = \int d\tau' \;
\begin{pmatrix} G_{LL}(\tau,\tau') & G_{LR}(\tau,\tau') \cr G_{RL}(\tau,\tau') &
G_{RR}(\tau,\tau') \cr\end{pmatrix} \begin{pmatrix} Q_L(\tau') \cr Q_R(\tau')\cr\end{pmatrix}
\end{equation}
where the index cq stands for coupling and source. As explained in section I,
we could also have set up a Lippmann-Schwinger equation for $\Psi^{cq}$ as
\begin{eqnarray}
\begin{pmatrix} \Psi^{cq}_L(\tau) \cr \Psi^{cq}_R(\tau) \cr\end{pmatrix} &=&
\int d\tau' \; \begin{pmatrix} g_{LL}(\tau,\tau') & 0 \cr 0 & g_{RR}(\tau,\tau')\end{pmatrix}
\begin{pmatrix} Q_L(\tau') \cr Q_R(\tau')\cr\end{pmatrix} + \nonumber \\ && \int d\tau' \;
\begin{pmatrix} g_{LL}(\tau,\tau') & 0 \cr 0 & g_{RR}(\tau,\tau')\end{pmatrix}
\begin{pmatrix} 0 & H_{LR} \cr H_{RL} & 0 \cr\end{pmatrix} \begin{pmatrix} \Psi^{cq}_L(\tau') \cr
\Psi^{cq}_R(\tau')\end{pmatrix} \nonumber \\
\end{eqnarray}
Now from inserting (10) into (11) and for $Q$ choosing some $\delta(\tau'-
\tau_0)$ either in the L- or in the R-component the Dyson equation for $G$
is obtained:
\begin{eqnarray}
&&\begin{pmatrix} G_{LL}(\tau,\tau') & G_{LR}(\tau,\tau') \cr
G_{RL}(\tau,\tau') & G_{RR}(\tau,\tau') \cr\end{pmatrix}=
\begin{pmatrix} g_{LL}(\tau,\tau') & 0 \cr 0 & g_{RR}(\tau,\tau') \cr\end{pmatrix}
+ \nonumber \\ && \int d\tau'' \;
\begin{pmatrix} g_{LL}(\tau,\tau'') & 0 \cr 0 & g_{RR}(\tau,\tau'') \cr\end{pmatrix}
\underbrace{
\begin{pmatrix} 0 & H_{LR} \cr H_{RL} & 0\cr\end{pmatrix}}_{\sigma}
\begin{pmatrix} G_{LL}(\tau'',\tau') & G_{LR}(\tau'',\tau') \cr
G_{RL}(\tau'',\tau') & G_{RR}(\tau'',\tau') \cr\end{pmatrix} \nonumber \\ 
\end{eqnarray}

\subsection{Explicit Green's functions in time and frequency domain}
\begin{figure}
\includegraphics[width=5cm]{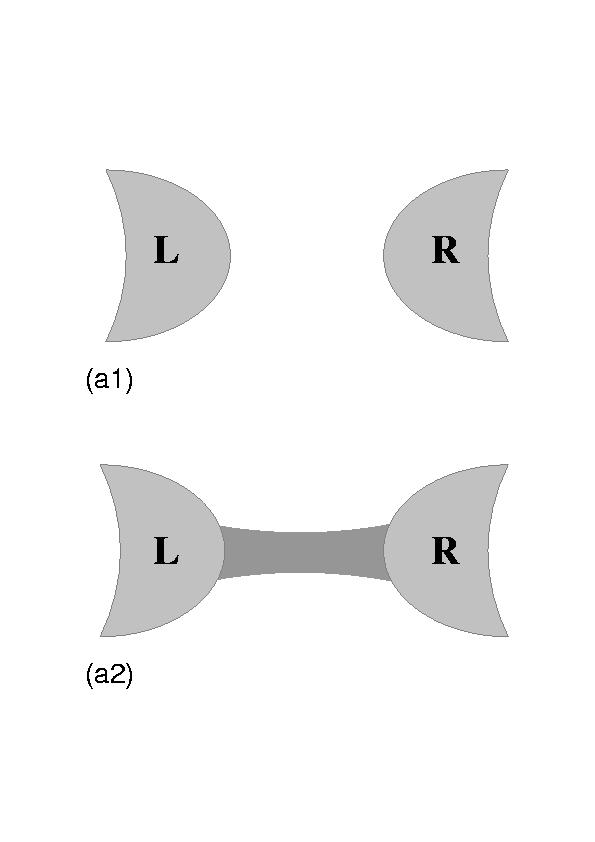}
\includegraphics[width=5.5cm]{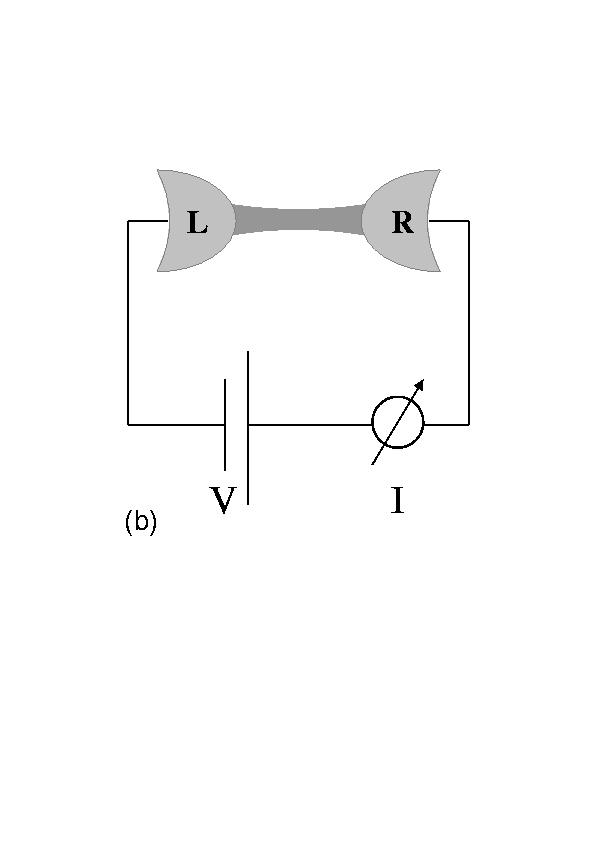}
\includegraphics[width=5.5cm]{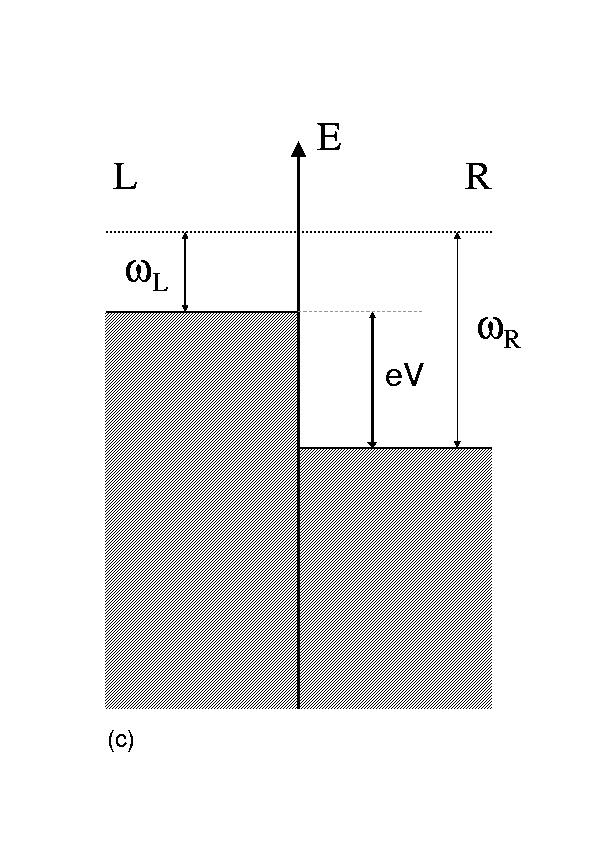}
\caption{A left and a right bulk reservoir, uncoupled (a1) and coupled (a2).
(b) Contact as embedded in a circuit.
(c) Fermi levels left and right with applied voltage. The dotted line
marks a fixed energy over both sides.} 
\end{figure}

The reservoirs on the left and right being bulk metal, the differential operator
of the homogeneous differential equation (1) is given by the free-partical
Hamiltonian
\begin{equation}
i\hbar\partial_{\tau}-H^0=i\hbar\partial_{\tau}+\frac{\hbar^2}{2m}\Delta
\end{equation}
and the corresponding wave functions are $\Psi^0=e^{\pm i\vec{k}\vec{r}-
i\omega \tau}$ according to the dispersion relation $E=\hbar\omega=
\frac{\hbar^2}{2m}\vec{k}^2$. Looking for a Green's function at given
frequency $\omega$, in (2) we can replace $H^0$ by $\hbar\omega$ and thus
have to solve
\begin{equation}
(i\hbar\partial_{\tau}-\hbar\omega)\; g(\tau,\tau',\omega)=\delta(\tau-\tau')
\end{equation}
We need an expression, the derivative of which produces $\delta(\tau-\tau')$.
One easily verifies that (14) is satisfied by either
\renewcommand{\theequation}{15\alph{equation}}
\setcounter{equation}{0}
\begin{eqnarray}
g^r(\tau,\tau',\omega)&=&-\frac{i}{\hbar}\; \theta(\tau-\tau')\; e^{-i\omega
(\tau-\tau')} \\ g^a(\tau,\tau',\omega)&=& \frac{i}{\hbar}\;
\theta(\tau'-\tau)\; e^{-i\omega (\tau-\tau')}
\end{eqnarray}
The retarded function $g^r$ only exists for $\tau \ge \tau'$ and the advanced
function $g^a$ for $\tau \le \tau'$. In (14) for simplicity we restricted
ourselves to a single energy and therefore (15) still contain $\omega$ as
a parameter. For complete Green's functions in the time domain these terms
have to be multiplied by the density of states $\mathcal{D} (\omega)$ and integrated
over energy. Later in transport we shall be interested in the amount of charge
transferred, not resolving any more, which energy levels contributions came
from. Thinking physically of a small contact between two metallic leads one might
argue that in the constriction transverse $k$-vectors are quantized and a
one-dimensional continuous density of states remains. Nevertheless, for not
too high voltages only a certain energy range around the Fermi energy will play
a role in transport and thus setting the density of states equal to its value
at the Fermi energy is a good approximation. Anyway, a constant density of
states $\mathcal{D}(\omega)=\mathcal{D}$ with occupied states below and empty ones
above the Fermi level for each side, left and right, shall just enter our
model here as an assumption. We are led to the following representations of
the retarded and advanced Green's functions:
\setcounter{equation}{0}
\renewcommand{\theequation}{16\alph{equation}}
\begin{eqnarray}
g^r(\tau,\tau')\; &=&-\frac{i}{\hbar}\; \theta(\tau-\tau') \int d\omega \;
\frac{\mathcal{D}(\omega)}{2\pi} \; e^{-i\omega(\tau-\tau')} \nonumber \\
&\widehat{=}& \frac{\mathcal{D}}{
h} \int d\omega \; g^r(\omega)\; e^{-i\omega(\tau-\tau')}  \quad
{\rm with} \quad g^r(\omega)=-i \\
g^a(\tau,\tau') \; &\widehat{=}& \frac{\mathcal{D}}{
h} \int d\omega \; g^a(\omega)\; e^{-i\omega(\tau-\tau')}  \quad
{\rm with} \quad g^a(\omega)=i 
\end{eqnarray}
\setcounter{equation}{16}
\renewcommand{\theequation}{\arabic{equation}}
Taking out the factor $\mathcal{D}/h$, we get the dimensionless functions
$g^r(\omega)=-i$ and $g^a(\omega)=i$ in frequency space. Remark that their
deduction here did not consist in calculating a Fourier transformation. The
phase factor $e^{-i\omega(\tau-\tau')}$ was already there in (15).
The $\omega$-integrals from (16) will be implemented as a useful representation
of $g^{r/a}$ still in the time domain.
The $\theta$-functions have been deliberately skipped after the
$\widehat{=}$-signs. We shall however see that $g^r$ and $g^a$ finally only
appear with the correct relation between their first and second time argument.
The $\omega$-integral should indeed be understood as summing over all energies
and, even if $g^{r/a}(\omega)$ is a constant, on no account be interpreted
as this constant times $2\pi\delta(\tau-\tau')$.
$e^{-i\omega(\tau-\tau')}=e^{-i\omega\tau} \cdot (e^{-i\omega\tau'})^*
=\vert\Psi^0(\tau)><\Psi^0(\tau')\vert$ is the
projection-like conversion of the phase unit vector
of an oscillation from time $\tau'$ to time $\tau$.
Left(L) and right(R) are distinguished as the origins of
wave functions constituting our basis of states, but the junction is considered
point-like, that is $\vec{r}=0$ for both left and right and thus $e^{\pm i
\vec{k}\vec{r}}=1$ drops out in an alike combination of $\Psi^0s$ even from
L and R. The two-fold time argument suggests to take (16) formally even as
double Fourier transform (however with different signs in the exponentials with
$\tau$ and $\tau'$)
\begin{equation}
\frac{h}{\mathcal{D}} \; g^{r/a}(\tau,\tau')=\int d\omega_1
\; \int d\omega_2 \; g^{r/a}(\omega_1,\omega_2) \; e^{-i\omega_1\tau} \;
e^{i\omega_2\tau'}
\end{equation}
with $g^{r/a}(\omega_1,\omega_2)=\mp i \delta(\omega_1-\omega_2)$, however,
because different energies stay independent. (More or less guessing the Green's
functions $g$ was easy in our normal-conducting simplest model here. Generally,
it has to be found as the solution to an equation like (14) with the differential
operator from the uncoupled system's Schr\"odinger equation and an elementary
perturbation $\delta$. In the superconducting state, for example, the
two-particle interaction in the Hamiltonian forces the Green's function to
include Andreev reflection \cite{ketterson}, and the result is such that $g(\omega)$ does not
just vanish in the gap of the quasi-particle density of states.)

Before we can solve (12) for the coupled system's Green's function $G$, we
have to specify the coupling parts of the Hamiltonian $H_{LR}$ and $H_{RL}$.
Like with the density of states the simplest model will assume that the
coupling is energy-independent and described by a constant (real) interaction
energy $W$. The reservoirs are unaltered by transport between them. Due to the
applied voltage (Fig.1b) incoming charge carriers (from the left on the right) are led
away and outgoing ones (on the left to the right) get replaced. In the contact
we do not allow relaxation or other energy-changing processes. We choose the
repective Fermi levels as zeros of energy on either side (Fig.1c). An
electron going from right to left has to strip off its phase $e^{-i\omega_R\tau}$
and aquire $e^{-i\omega_L\tau}$ to fit in on the left. An analogous argument
holds for transitions from left to right, and therefore the coupling terms are
\setcounter{equation}{0}
\renewcommand{\theequation}{18\alph{equation}}
\begin{eqnarray}
H_{LR}&=&\sigma_{LR}(\tau)=W\; e^{-i(\omega_L-\omega_R)\tau}=W\; e^{ieV\tau/\hbar} \\
H_{RL}&=&\sigma_{RL}(\tau)=W\; e^{-i(\omega_R-\omega_L)\tau}=W\; e^{-ieV\tau/\hbar}
\end{eqnarray}
\setcounter{equation}{18}
\renewcommand{\theequation}{\arabic{equation}}
The phase factor gives a time-dependence to $H_{LR}$ and $H_{RL}$, but
$\omega_R-\omega_L=eV/\hbar$ is indeed independent of energy. For $G$ we make
an ansatz like (17) as a two-fold Fourier representation:
\begin{equation}
\frac{h}{\mathcal{D}} \; G(\tau,\tau') = \int d\omega_1 \;
\int d\omega_2 \; G(\omega_1,\omega_2) \; e^{-i\omega_1\tau} \;
e^{i\omega_2\tau'}
\end{equation}
From (12), which is valid for either advanced or retarded functions, as an
example, we pick the upper left component of the 2x2 matrix in LR-space and
insert (17), (18) and (19): (Multiple integral signs are skipped from now on.)
\begin{eqnarray}
G_{LL}(\tau,\tau')&=&g_{LL}(\tau,\tau')+\int d\tau'' \; g_{LL}(\tau,\tau'') \;
H_{LR}(\tau'') \; G_{RL}(\tau'',\tau') \quad \Longleftrightarrow \nonumber \\ &&
\frac{\mathcal{D}}{h} \int d\omega_{L1} \; d\omega_{L2} \;
G_{LL}(\omega_{L1},\omega_{L2}) \; e^{-i\omega_{L1}\tau} \;
e^{i\omega_{L2}\tau'} = \nonumber \\ &&
\frac{\mathcal{D}}{h} \int d\omega_{L1} \; d\omega_{L2} \;
g_{LL}(\omega_{L1},\omega_{L2}) \; e^{-i\omega_{L1}\tau}\;
e^{i\omega_{L2}\tau'} + \nonumber \\ &&
\int d\tau'' \; d\omega_{L1} \; d\omega_{L3} \; d\omega_{R} \; d\omega_{L2} \;
\frac{\mathcal{D}}{h} \; g_{LL}(\omega_{L1},\omega_{L3})\;
e^{-i\omega_{L1}\tau} \; e^{i\omega_{L3}\tau''}\; \cdot \nonumber \\ &&
\cdot \; W \; e^{ieV\tau''/\hbar}\;
\frac{\mathcal{D}}{h}\; G_{RL}(\omega_R,\omega_{L2}) \;
e^{-i\omega_R\tau''}\; e^{i\omega_{L2}\tau'}
\end{eqnarray}
The integral over $\tau''$ produces $2\pi\delta(\omega_{L3}+\frac{eV}{\hbar}-
\omega_R)$ and in $g(\omega_{L1},\omega_{L3})$ there is $\delta(\omega_{L1}-
\omega_{L3})$ anyway, such that the last term of (20) becomes 
$$\frac{\mathcal{D}}{h} \int d\omega_{L1} \; d\omega_{L2} \; g(\omega_{L1}) \; t \;
G_{RL}(\omega_{L1}+\frac{eV}{\hbar},\omega_{L2}) \; e^{-i\omega_{L1} \tau} \;
e^{i\omega_{L2} \tau'} $$
with $t=W\mathcal{D}/\hbar$.
Strictly speaking, if (20) were for the retarded function, in the last term
there would be $\theta(\tau-\tau'')$ from $g^r_{LL}$ and $\theta(\tau''-\tau')$
from $G^r_{RL}$, and if it were for the advanced function, $\theta(\tau''-\tau)$
and $\theta(\tau'-\tau'')$, such that the integral over $\tau''$ only exists
between $\tau$ and $\tau'$ instead of having minus and plus infinity as limits.
However, with $\omega_{L3}$ and $\omega_R$ running over any value, one can argue
that even with a finite $\tau''$-integral the only remaining contribution
stems from $\omega_{L3}+\frac{eV}{\hbar}-\omega_R=0$. A discussion of time
ordering will again appear in section 5.
$\mathcal{D}$ is the density of states per frequency interval, dividing by $\hbar$
makes it the number of states per energy interval. $W$ is an energy. $t$ can
be understood as a dimensionless transmission amplitude. Now we set up the
convention that all frequency arguments of $g$ and $G$ are written with respect
to the left zero level and for the case that they correspond to the right,
that is an R-index, it is understood that $eV/\hbar$ is added. (20) has
to hold for any $\tau$ and $\tau'$ and from comparing Fourier coefficients we get
\setcounter{equation}{19}
\renewcommand{\theequation}{\arabic{equation}a}
\begin{equation}
G_{LL}(\omega_{L1},\omega_{L2})=g_{LL}(\omega_{L1}) \; \delta(\omega_{L1}-
\omega_{L2}) + 
g_{LL}(\omega_{L1}) \; t \; G_{RL}(\omega_{L1},\omega_{L2})
\end{equation}
We could have inserted $G_{RL}=g_{RR}H_{RL}G_{LL}$ and again replaced
$G_{LL}=g_{LL}+g_{LL}H_{LR}G_{RL}$ and so on. Instead of an implicit equation
for $G$ this would have led to an infinite series (see section I):
\setcounter{equation}{20}
\renewcommand{\theequation}{\arabic{equation}}
\begin{equation}
G=g\sum_{n=0}^{\infty} (\sigma g)^n=\left( \sum_{n=0}^{\infty} (g\sigma)^n
\right) g =
g+g\sigma g+g\sigma g \sigma g + \ldots
\end{equation}
(21) is written for whole matrices in LR-space, calculations like (20) can be
done analogously for $G_{LR}$, $G_{RL}$ and $G_{RR}$. (21) can be
read as equation in the time domain. Then $\sigma$ is the matrix consisting of
$H_{LR}$ and $H_{RL}$ and each multiplication of two following $g$s with
$\sigma$ in between means an integration over time. However, (21) is as well
valid as relation in frequency space. In this case $\sigma$ is the 2x2 matrix
with just $t$ as off-diagonal elements. Like we have seen through evaluating
the $\tau''$-integral in (20), in (21) each connection $H_{LR/RL}$ from
$\sigma$ passes the frequency argument from the $g$ in front to the $g$ behind.
And as $g$ can only have two identical frequency arguments, no $\omega$
different from the first can ever appear, such that $\omega_{L1}=\omega_{L2}=
\omega$ and $G$ also effectively is a function of only one frequency argument:
$G(\omega_1,\omega_2)=\delta(\omega_1-\omega_2)G(\omega_1)$. With
Green's functions of a single frequency argument (20a) and its analogues for the
other three components in LR-space become a simple algebraic equation:
\begin{equation}
\begin{pmatrix} G_{LL}(\omega) & G_{LR}(\omega) \cr G_{RL}(\omega) & G_{RR}(\omega)\cr\end{pmatrix}=
\begin{pmatrix} g_{LL}(\omega) & 0 \cr 0 & g_{RR}(\omega) \cr \end{pmatrix}
+\begin{pmatrix} g_{LL}(\omega) & 0 \cr 0 & g_{RR}(\omega) \cr \end{pmatrix}
\begin{pmatrix} 0 & t\cr t & 0\cr\end{pmatrix}
\begin{pmatrix} G_{LL}(\omega) & G_{LR}(\omega) \cr G_{RL}(\omega) & G_{RR}(\omega)\cr\end{pmatrix}
\end{equation}
(The fact that even the Green's function of the coupled system turns out to
be a function of a single frequency argument is a special feature of our
simple model for the normal conducting case. In the extension of this model
to superconducting reservoirs and transmission processes including Andreev
reflection, $G$ becomes a function of two frequency arguments, the second,
however, restricted to values differing from the first by an integer multiple
of $eV/\hbar$, such that effectively there is one continuous and one discrete
frequency parameter \cite{Cuequ}.) Here, with $g^{r/a}(\omega)=\mp i$, (22) is easily
solved and $G(\omega)$ comes out independent of frequency, too:
\begin{eqnarray}
\begin{pmatrix} G_{LL} & G_{LR} \cr G_{RL} & G_{RR} \cr\end{pmatrix}^{r/a}(\omega)&=&
\left[ \begin{pmatrix} 1 & 0 \cr 0 & 1 \cr\end{pmatrix} -
\begin{pmatrix} \mp i & 0 \cr 0 & \mp i \cr\end{pmatrix}
\begin{pmatrix} 0 & t \cr t & 0 \cr\end{pmatrix} \right]^{-1}
\begin{pmatrix} \mp i & 0 \cr 0 & \mp i \cr\end{pmatrix}
\nonumber \\
&=&\frac{1}{1+t^2}\begin{pmatrix} \mp i & -t \cr -t & \mp i \cr\end{pmatrix}
\end{eqnarray}
The same result could have been obtained from (21) by writing out a few more
of the matrix multiplications and using the formula for the geometric series
in each element. We shall need two further types of Green's functions. $T$
will be introduced in the next section and $g^{+-}$ and $G^{+-}$ when
calculating the current.

\subsection{Transfer Green's functions}
In (21) there was a sum of products of arbitrary many factors $g$ and $\sigma$
with outer factors $g$. ''Product'', of course, except with the Green's functions
taken of a single frequency parameter, in the time domain or with two-fold
frequency dependence still meant a convolution-type integration over inner
arguments. In analogy we define the sum of products with outer factors
$\sigma$ (as integrals in the time domain or just algebraically with
$g(\omega)$ and $\sigma=\begin{pmatrix} 0 & t \cr t & 0\cr\end{pmatrix}$):
\begin{equation}
T=\sigma+\sigma g \sigma +\sigma g\sigma g\sigma + \ldots =
\sigma \sum_{n=0}^{\infty}(g\sigma)^n=\left( \sum_{n=0}^{\infty} (\sigma g)^n
\right) \sigma
\end{equation}
Whereas all contributions to $G$ in (21) began and ended with staying some time
in a reservoir, described by $g$ - at L from $\tau$ to $\tau''$ for a start
in the last term of (20), for example - each term of $T$ in (24) begins and
ends with a transition $\sigma$ and further contains at least one such hopping
across the junction (which $g$ does not). Therefore we call $T$ the {\it
transfer} Green's function.
The same as for $G$, the relation between $T$ as a function of times and as
a function of frequency is given by
\begin{equation}
\frac{\mathcal{D}}{\hbar}\; T_{JK}(\tau,\tau')=\frac{1}{2\pi}\int d\omega \;
T_{JK}(\omega)\; e^{-i\omega_J\tau} \; e^{i\omega_K\tau}
\end{equation}
where $\omega_J=\omega$ if J=L and $\omega_J=\omega+eV/\hbar$ if J=R and the
same for $\omega_K$.
(Taking out $2\pi$ of $\mathcal{D}$ in (16) was a convention.
The prefactor of the $\omega$-integral for $T$ follows from consistency
requirements. (19) for $G$ was in complete analogy to (16) for $g$.
$T$ with two time arguments, however, has a little different character from
$\sigma$ with just one time parameter.)
Alternatively to (24) $T$ could be defined through its link to $G$
\begin{equation}
\sigma G=Tg \quad {\rm or} \quad G\sigma =gT
\end{equation}
Be careful that replacing one by the other can introduce another internal
time integration as, for example,
$\sigma(\tau)\; G(\tau,\tau')=\int d\tau''\; T(\tau,\tau'')\; g(\tau'',\tau')$.
(24) and (26) hold for retarded and advanced functions.
From (24) it is immediately seen that $T$ like $G$
satisfies a Dyson equation
\begin{equation}
G=g+g\sigma G \quad {\rm and} \quad T=\sigma+\sigma g T
\end{equation} 
and even the complementary forms
\begin{equation}
G=g+G\sigma g \quad {\rm and} \quad T=\sigma+Tg\sigma
\end{equation}
are analogues. In Fourier space, like (22) the $T$-equation (27) is an
algebraic equation and the solution like
\begin{equation}
G=({\bf 1}-g\sigma )^{-1}g \quad {\rm is} \quad T=({\bf 1}-\sigma g)^{-1}\sigma.
\end{equation}
Inserting $g$ and $\sigma$ explicitly for our model we get
\begin{eqnarray}
\begin{pmatrix} T_{LL} & T_{LR}\cr T_{RL} & T_{RR}\cr\end{pmatrix}^{r/a}(\omega)&=&
\left[ \begin{pmatrix} 1 & 0\cr 0 & 1\cr\end{pmatrix}-\begin{pmatrix}0 & t\cr t & 0\cr\end{pmatrix}
\begin{pmatrix} \mp i & 0 \cr 0 & \mp i\cr\end{pmatrix} \right]^{-1}
\begin{pmatrix} 0 & t\cr t & 0\cr\end{pmatrix} \nonumber\\
&=& \frac{1}{1+t^2}\begin{pmatrix} \mp it^2 & t \cr t & \mp it^2\cr\end{pmatrix}
\end{eqnarray}
especially
\begin{equation}
T_{LR}^{r/a}=T_{RL}^{r/a}=\frac{t}{1+t^2}=t-t^3+t^5-\ldots
\end{equation}

\begin{figure} \begin{center}
\includegraphics[width=5cm]{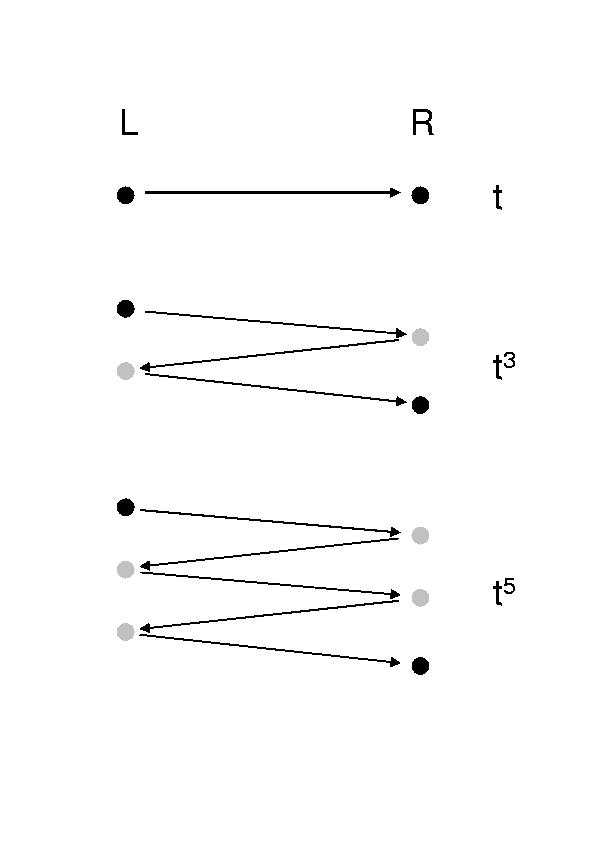}  \end{center}
\caption{Transfer processes from left to right of different order.}
\end{figure}

Whereas $t$ is the single hopping amplitude, $T_{LR/RL}$ is a renormalized
transfer amplitude. One may wonder why a model for transport could not have
been set up adding amplitudes for transfer processes of all orders, as the
interaction (18) seems to be introduced the way it is just in order to result
in powers of $t$. However, $g^{r/a}(\omega)=\mp i$ deduced from the
Schr\"odinger equation is decisive for the signs in (30) and (31). One may
wonder that multiple reflections are not added as $t+t^3+t^5+\ldots=
\frac{t}{1-t^2}$. Processes of different order (Fig.2) are not independent,
but interfere. $T_{RL}$ is the transfer amplitude per single electron supplied
on the left by the voltage source. But for every electron that goes over to
the right in an $n$th order process ($n$ odd)
with weight $t^2$ there is one that has hopped
once more to the right and back ($n+2$ order process) and thus is not to be
newly supplied, but to be again sent through the junction.
The amplitude $t$ is renormalized by $1+t^2$ as denominator.
However, such interpretations of quantum mechanical amplitudes are precarious,
and the full conversion of $t$ to a transmission probability will be
established later.

It is quite instructive to solve (27) in a slightly different way than done in
(30). Firstly, for the four components in LR-space we have
\renewcommand{\theequation}{\arabic{equation}a}
\setcounter{equation}{26}
\begin{eqnarray}
T_{LL}=\phantom{\sigma_{RL}+} \sigma_{LR}g_{RR}T_{RL} \qquad
T_{LR}=\sigma_{LR}+\sigma_{LR}g_{RR}T_{RR} \nonumber \\
T_{RL}=\sigma_{RL}+\sigma_{RL}g_{LL}T_{LL} \qquad
T_{RR}=\phantom{\sigma_{RL}+} \sigma_{RL}g_{LL}T_{LR}
\end{eqnarray}
\renewcommand{\theequation}{\arabic{equation}}
\setcounter{equation}{31}
Inserting these into each other, for example, an equation for $T_{LR}$ alone
is obtained:
\begin{equation}
T_{LR}=\sigma_{LR}+\sigma_{LR}g_{RR}\sigma_{RL}g_{LL}T_{LR}
\end{equation}
This implicit equation is the basis for calulating the transfer Green's
function in more complicated cases than discussed here
\cite{Cuequ,my06}, like for example
the superconducting junction. In our model, inserting $\sigma_{LR}=
\sigma_{RL}=t$ and $g^{r/a}_{LL}(\omega)=g^{r/a}_{RR}(\omega)=\mp i$ into (32)
immediately also leads to $T_{LR}^{r/a}(\omega)=\frac{t}{1+t^2}$.

\begin{figure} \begin{center}
\includegraphics[width=7cm,angle=270]{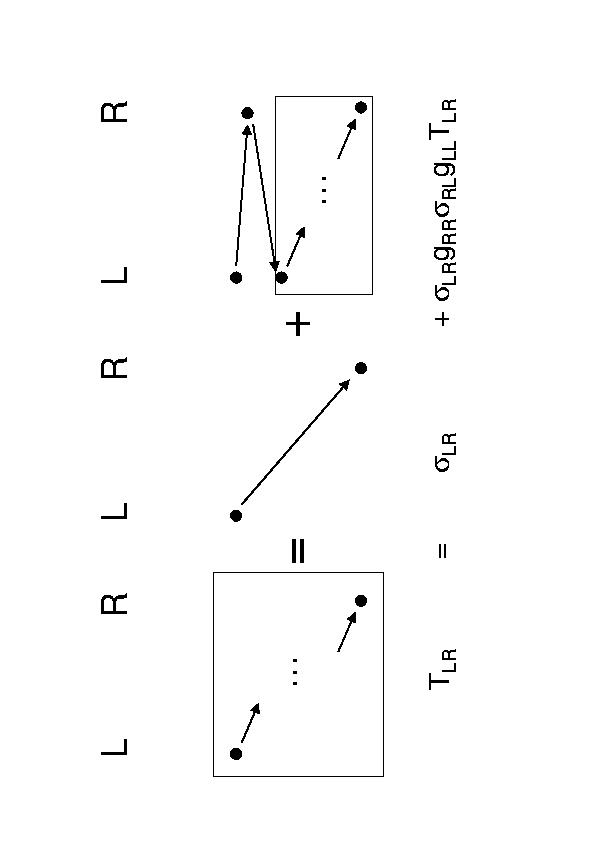} \end{center}
\caption{Illustration of the Dyson equation for the transfer Green's function.}
\end{figure}

More easily than the Dyson equation for the ordinary Green's function
$G$, the one for the transfer Green's function $T$ is illustrated as is
done for the LR-component in Fig.3. (Normally indices are read from right
to left such that $T_{LR}$ is considered a transition from right to left,
but it does not really matter whether they are interpreted the other way
round as in Fig.3. The actual sequence of what is earlier or later in time
will be discussed when calculating the current in the next section.)
Fig.3 demonstrates the implicitness of the Dyson equation: Any transition
from left to right is either a single transfer or an electron hopping to the
right and back followed by any process beginning on the left and ending on
the right, no matter what happens in between. This last part by definition
is the sane as the other side of the equation, namely $T_{LR}$. 

\subsection{Calculating the current}
From the Heisenberg picture of quantum mechanics we know that the time derivative
of a not explicitly time-dependent operator $A$ is given by the commutator
with the Hamiltonian \cite{CoTan_2}:
\begin{equation}
\frac{d}{d\tau}A=\frac{i}{\hbar}[H,A]
\end{equation}
The operator of interest here is the projector on either side of the junction
\begin{equation}
\rho_L=\frac{\vert \Psi^0_L><\Psi^0_L\vert}{\vert \Psi^0_L \vert^2} \widehat{=}
\begin{pmatrix} 1 & 0\cr 0 & 0\cr\end{pmatrix} \quad {\rm or} \quad
\rho_R=\frac{\vert \Psi^0_R><\Psi^0_R\vert}{\vert \Psi^0_R \vert^2} \widehat{=}
\begin{pmatrix} 0 & 0\cr 0 & 1\cr\end{pmatrix}
\end{equation}
As explained earlier, with the junction coupling left and right together, the
solution $\begin{pmatrix}\Psi_L \cr \Psi_R \cr\end{pmatrix}$ is not limited to one side, however,
the projectors take out the respective part:
\setcounter{equation}{0}
\renewcommand{\theequation}{34\alph{equation}}
\begin{equation}
\begin{pmatrix} 1 & 0\cr 0 & 0\cr\end{pmatrix}
\begin{pmatrix} \Psi_L \cr \Psi_R \cr\end{pmatrix}=
\begin{pmatrix} \Psi_L \cr 0 \cr\end{pmatrix} \quad {\rm and} \quad
\begin{pmatrix} 0 & 0\cr 0 & 1\cr\end{pmatrix} \begin{pmatrix}\Psi_L \cr \Psi_R \cr\end{pmatrix}=
\begin{pmatrix} 0 \cr \Psi_R\cr\end{pmatrix}
\end{equation}
\renewcommand{\theequation}{\arabic{equation}}
\setcounter{equation}{34}
$\rho_L$ and $\rho_R$ are proportional to the amount of charge on the left and
on the right side. Their time derivatives have equal absolute values, but
opposite sign and represent the current.
\begin{equation}
I=-e<\frac{d\rho_L}{d\tau}>=e<\frac{d\rho_R}{d\tau}>
\end{equation}
($<\; \vert$ and $\vert\; >$ are used for {\tt bra}- and {\tt ket}-states. Here $<\; >$ means
the expectation value, of course.)
$e$ is the charge of an electron and the sign of $I$ can be defined
arbitrarily. We choose to do the calculation with $\rho_L$ and evaluate
the commutator with the Hamiltonian:
\begin{equation}
[H,\rho_L]=\left[ \begin{pmatrix} H_{LL} & H_{LR} \cr H_{RL} & H_{RR} \cr\end{pmatrix},
\begin{pmatrix} 1 & 0 \cr 0 & 0 \cr\end{pmatrix} \right] = \underbrace{
\begin{pmatrix} 0 & -H_{LR} \cr H_{RL} & 0 \cr\end{pmatrix}}_{\sigma_c}
\end{equation}
Putting together (33) and (35) we obviously need the expectation value of
the operator $[H,\rho_L]$. $[H,\rho_L]$ shall be called $\sigma_c$.
Even if (33) stems from the Heisenberg picture, it is written in such a way,
that the right hand side is to be evaluated in the Schr\"odinger system with
time dependent states, and we shall here change to the interaction picture
\cite{Mahan}
for the calculation. With $^{\dagger}$ standing for complex conjugation
as well as transposition from column to line vector, the value of
${\displaystyle \frac{d\rho_L}{dt}}$ in state $\begin{pmatrix}
\Psi_L\cr \Psi_R\cr\end{pmatrix} $ at time $\tau$ is given by
\begin{eqnarray} &&
\begin{pmatrix} \Psi_L(\tau)\cr \Psi_R(\tau)\end{pmatrix}^{\dagger} \;
\frac{i}{\hbar} \; \begin{pmatrix} 0 &
-H_{LR}(\tau) \cr H_{RL}(\tau) & 0 \end{pmatrix} \;
\begin{pmatrix} \Psi_L(\tau) \cr
\Psi_R(\tau)\end{pmatrix} = \nonumber \\ &&
<\Psi^0(\tau_0) \vert \widetilde{T} \; \exp (\frac{1}{\hbar}\int_{\tau_0}^{\tau} i \; \sigma_{
\mathcal{H}}(\tau') \; d\tau') \; i[H,\rho_L]_{\mathcal{H}} \; \widehat{T} \;
\exp (\frac{1}{\hbar}\int_{\tau_0}^{\tau} (-i) \; \sigma_{\mathcal{H}}(\tau') \; d\tau') \vert
\Psi^0(\tau_0)> \nonumber \\
\end{eqnarray}
$H_{LR}(\tau)$ and $H_{RL}(\tau)$ denote the time occurence of $\tau$ from (18)
which has to be considered as still belonging to the Schr\"odinger picture
in our case here. $\widehat{T}$ means time ordering \cite{Mahan} and $\widetilde{T}$
anti-time ordering. $\Psi^0$ in the uncoupled system, of course, also stands
for a two-vector with left and right component.
Replacing $\begin{pmatrix} \Psi_L(\tau)\cr \Psi_R(\tau)\end{pmatrix}$ by $\Psi^0(\tau_0)$ in (37) we took out both
the coupling as well as the time dependence from the states. Accordingly the
$\sigma_{\mathcal{H}}$ in the integrals, in contrast to the
Schr\"odinger-picture $\sigma$ used in the preceding sections, in a
Heisenberg way have to include the time dependence of the uncoupled states.
$\sigma(\tau)$ could already be written as
$$ \sigma(\tau)=\begin{pmatrix} 0 & H_{LR}(\tau) \cr H_{RL}(\tau) & 0 \cr\end{pmatrix}=
\begin{pmatrix} e^{-i\omega_L\tau} & 0\cr 0 & e^{-i\omega_R\tau}\cr\end{pmatrix}
\begin{pmatrix} 0 & W\cr W & 0\cr\end{pmatrix}
\begin{pmatrix} e^{i\omega_L\tau} & 0\cr 0 & e^{i\omega_R\tau}\cr\end{pmatrix}$$
or $\sigma(\tau)=\vert\Psi^0(\tau)>W<\Psi^0(\tau)\vert$ for short, where
{\tt bra}, {\tt ket} and $W$ still mean the respective matrices. However,
this decomposition might rather be confusing and will not be used, anyway.
The translation to the interaction picture is the following:
\begin{eqnarray}
\sigma_{\mathcal{H}}&=&\begin{pmatrix} e^{\frac{i}{\hbar}H_{LL}(\tau-\tau_0)} & 0 \cr 0 &
e^{\frac{i}{\hbar}H_{RR}(\tau-\tau_0)}\cr\end{pmatrix} \begin{pmatrix}
0 & H_{LR}(\tau) \cr H_{RL}(\tau) & 0 \cr\end{pmatrix}
\begin{pmatrix} e^{-\frac{i}{\hbar}H_{LL}(\tau-\tau_0)} & 0 \cr 0 &
e^{-\frac{i}{\hbar}H_{RR}(\tau-\tau_0)}\cr\end{pmatrix}\nonumber \\ &=&
\begin{pmatrix} e^{-i\omega_L\tau_0} & 0 \cr 0 & e^{-i\omega_R\tau_0}\cr\end{pmatrix}
\begin{pmatrix} e^{i\omega_L\tau} & 0 \cr 0 & e^{i\omega_R\tau}\cr\end{pmatrix} \; 
\begin{pmatrix} 0 & H_{LR}(\tau) \cr H_{RL}(\tau) & 0 \cr\end{pmatrix} \cdot \nonumber \\ &&
\qquad \qquad \qquad \qquad \qquad \qquad
\begin{pmatrix} e^{-i\omega_L\tau} & 0 \cr 0 & e^{-i\omega_R\tau}\cr\end{pmatrix}
\begin{pmatrix} e^{i\omega_L\tau_0} & 0 \cr 0 & e^{i\omega_R\tau_0}\cr\end{pmatrix}
\end{eqnarray}
or $\sigma_{\mathcal{H}}(\tau)=\vert\Psi^0(\tau_0)><\Psi^0(\tau)\vert\;
\sigma(\tau)\; \vert\Psi^0(\tau)><\Psi(\tau_0)\vert$. The Heisenberg picture
always refers the operator back to the undeveloped state at $\tau_0$.

\begin{figure}\begin{center}
\includegraphics[width=8cm,angle=270]{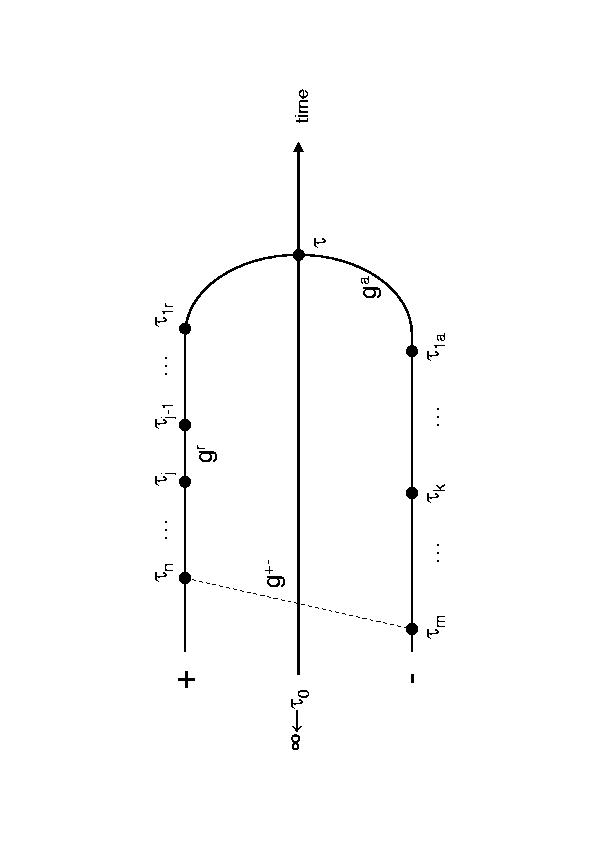}
\end{center}
\caption{The Keldysh contour illustrating the development
out of the uncoupled system's states of the {\tt bra} $<\Psi(\tau)\vert$ on
the minus and the {\tt ket} $\vert \Psi(\tau)>$ on the plus branch.
This view picks out the single transition at time $\tau$ and visualizes
the calculation of $<\Psi(\tau)\vert \sigma(\tau) \vert \Psi(\tau)>$.
Each point on the contour represents an LR- or RL-transition described by
$\sigma$, each line segment the development of the wave-function phase
which is given $g$.}
\end{figure}

Analogously $\sigma_c$ will have to be extended by the time-dependence of the
uncoupled states. 
The meaning of the time-integral over $\sigma_{\mathcal{H}}$ as an exponential
is best explained by writing explicitly:
\begin{eqnarray}
\widehat{T}\; \exp (\frac{1}{\hbar} \int_{\tau_0}^{\tau}(-i)\;
\sigma_{\mathcal{H}}(\tau')\; d\tau') = \sum_n
\frac{1}{\hbar} \int_{\tau_2}^{\tau} d\tau_1 \ldots
\frac{1}{\hbar} \int_{\tau_{j+1}}^{\tau_{j-1}} d\tau_j \ldots
\frac{1}{\hbar} \int_{\tau_0}^{\tau_{n-1}} d\tau_n \cdot \nonumber \\ 
(-i) \sigma_{\mathcal{H}}(\tau_1) \ldots (-i) \sigma_{\mathcal{H}}(\tau_j)
\ldots (-i) \sigma_{\mathcal{H}}(\tau_n)
\end{eqnarray}
where all arguments $\tau_1,\ldots ,\tau_j,\ldots ,\tau_n$ have to lie between
$\tau_0$ and $\tau$ and all products from none to arbitrarily many factors
$\sigma_{\mathcal{H}}$ have to be added. With time ordering, furthermore
$\tau_1\ge\ldots\ge\tau_j\ge\ldots\ge\tau_n$ is imposed. Without $\widehat{T}$
the scheme for the exponential series $e^x=\sum_{n=0}^{\infty}\frac{x^n}{n!}$
in (39) would produce $\sum_n \frac{1}{n!} \; \frac{1}{\hbar}
\int_{\tau_0}^{\tau}d\tau_1 \ldots
\frac{1}{\hbar} \int_{\tau_0}^{\tau}d\tau_j \ldots
\frac{1}{\hbar} \int_{\tau_0}^{\tau}d\tau_n \ldots$ .
But then for every fixed set $\{ \tau_1,\ldots ,\tau_j,\ldots ,\tau_n\}$
there are $n!$ permutations for the time values to occur such that time
ordering and restricting the arguments to mutually exclusive intervals as
done in (39) cancels the factorial denominators.
Writing out the anti-time ordered part in the same way as shown for the
time-ordered part, the expression from (37) becomes
\renewcommand{\theequation}{37\alph{equation}}
\setcounter{equation}{0}
\begin{eqnarray} && 
{\bf <\Psi^0(\tau_0)\vert} \quad \Psi^0(\tau_0)> \frac{i}{\hbar} <\Psi^0(\tau_m)\vert
\sigma(\tau_m)\vert \Psi^0(\tau_m)><\Psi^0(\tau_0)\vert \; \ldots
\nonumber \\ &&
\ldots \vert \Psi^0(\tau_0)> \frac{i}{\hbar} <\Psi^0(\tau_k)\vert\sigma(\tau_k)\vert
\Psi^0(\tau_k)><\Psi^0(\tau_0)\vert \quad
{\it \Psi^0(\tau_0)> \frac{i}{\hbar} <\Psi^0(\tau_{k-1})\vert}
\nonumber \\ &&
{\it \sigma(\tau_{k-1})\vert}\ldots
\vert\sigma(\tau_{1a})\vert\Psi^0(\tau_{1a})><\Psi^0(\tau_0)\vert
\quad {\bf \Psi^0(\tau_0)> \frac{i}{\hbar} <\Psi^0(\tau)\vert\sigma_c(\tau)\vert}
\nonumber \\ &&
{\bf \Psi^0(\tau)><\Psi^0(\tau_0)\vert}\quad
\Psi^0(\tau_0)>(\frac{-i}{\hbar})<\Psi^0(\tau_{1r})\vert\sigma(\tau_{1r})\vert \ldots
\nonumber \\ &&
{\it \ldots \vert\sigma(\tau_{j-1})\vert\Psi^0(\tau_{j-1})><\Psi^0(\tau_0)}
\quad \vert \Psi^0(\tau_0)>(\frac{-i}{\hbar})<\Psi^0(\tau_j)\vert\sigma(\tau_j)\vert
\Psi^0(\tau_j)>
\nonumber \\ &&
<\Psi^0(\tau_0)\vert
\ldots \vert\sigma(\tau_n)\vert\Psi^0(\tau_n)><\Psi^0(\tau_0) \quad
{\bf \vert \Psi^0(\tau_0)>}
\end{eqnarray}
where $\tau_0\le\tau_m\le\ldots\le\tau_k\le\ldots\tau_{1a}\le\tau$,
$\tau_0\le\tau_n\le\ldots\le\tau_j\le\ldots\le\tau_{1r}\le\tau$
and integration over all time arguments except $\tau$ and $\tau_0$ is
understood. Like in (39) we mean the sum of all products with arbitrary
many inner development factors like $\Psi^0$ of $\tau_k$ and
$\Psi^0$ of $\tau_j$. Spaces and different font types are just used in (37a)
to recognize sequential $\sigma_{\mathcal{H}}$-parts like (38).
{\it Inner} development factors like $\Psi^0$ of $\tau_k$ and
$\tau_j$ as well as $\Psi^0(\tau)$ and the inner $\Psi^0(\tau_0)$
have to sum over space or the basis of states and therefore should be
thought of as matrices as in (38).
We have not yet decided which components {\it outer}
{\tt bra} and {\tt ket} with ${\bf \Psi^0(\tau_0)}$ will project out.
All $<\Psi^0(\tau_0)\vert\Psi^0(\tau_0)>$ are the unit matrix and drop out
($\Psi^0$ being given by a non-normalizable $e$-function does not pose
a problem here).
Then $\vert\Psi^0(\tau_k)>\frac{i}{\hbar}<\Psi^0(\tau_{k-1})\vert$
with $\tau_k\le\tau_{k-1}$ is recognized as $g^a(\tau_k,\tau_{k-1})$ and
$\vert\Psi^0(\tau_{j-1})>\frac{(-i)}{\hbar}<\Psi^0(\tau_j)\vert$
with $\tau_{j-1}\ge\tau_j$ as $g^r(\tau_{j-1},\tau_j)$. These $g^{a/r}$
are 2x2 diagonal matrices.
With (21) in (37a) the whole sequence from $\vert\Psi^0(\tau_m)>$ to
$<\Psi^0(\tau)\vert$ can be replaced by $G^a(\tau_m,\tau)$ and the long part
from $\vert\Psi^0(\tau)>$ up to $<\Psi^0(\tau_n)\vert$ is just $G^r(\tau,\tau_n)$.
The complete operator between the outermost ${\bf <\Psi^0(\tau_0)}\vert$ and
${\bf \vert\Psi^0(\tau_0)>}$ in (37a) therefore becomes
\begin{eqnarray}
\vert\Psi^0(\tau_0)>\frac{i}{\hbar}<\Psi^0(\tau_m)\vert 
\left[ {\bf 1}\; \delta(\tau_m-\tau)+\sigma(\tau_m)G^a(\tau_m,\tau)\right]
\sigma_c(\tau) \nonumber \\
\left[ {\bf 1}\; \delta(\tau_n-\tau)+G^r(\tau,\tau_n)\sigma(\tau_n)\right]
\vert\Psi^0(\tau_n)><\Psi^0(\tau_0)\vert
\end{eqnarray}
The ${\bf 1}$-contributions stem from the cases where there are no
factors with $\Psi^0(\tau_k)$ or $\Psi^0(\tau_j)$.
We regard the coupled system as having developed out of the uncoupled system,
but we are looking for a stationary state. To achieve this, the coupling has
to have been turned on infinitely long ago, thus we let $\tau_0\rightarrow
-\infty$. The sequence of time arguments is usually represented on the
so-called Keldysh contor (Fig.4). We still have to take the expectation
value of (37) or (37a). This means summing over the basis of states for the
outer ${\bf \Psi^0(\tau_0)}$. The uncoupled basis consists of $\begin{pmatrix}
1\cr 0\cr\end{pmatrix}$
and $\begin{pmatrix} 0\cr 1\cr\end{pmatrix}$ for a state on the left and a state on the right
for each frequency $\omega$. We shall discuss the occupation or emptiness of
states shortly after having worked off some further more formal points.
Summing over $\begin{pmatrix} 1\cr 0\cr\end{pmatrix}$ and $\begin{pmatrix}
0\cr 1\cr\end{pmatrix}$
for the outer ${\bf \Psi^0(\tau_0)}$ will return the trace of the matrix $M$
given in (37b).
$$ \begin{pmatrix} 1\cr 0\cr\end{pmatrix}^{T} \underbrace{\begin{pmatrix}
m_{11} & m_{12}\cr m_{21} & m_{22}\cr\end{pmatrix}}_M \begin{pmatrix}
1\cr 0\cr\end{pmatrix} + \begin{pmatrix}0\cr 1\cr\end{pmatrix}^{T}
\underbrace{\begin{pmatrix} m_{11} & m_{12}\cr m_{21} & m_{22}\cr\end{pmatrix}}_M
\begin{pmatrix} 0\cr 1\cr\end{pmatrix} = {\rm Tr}\; M =m_{11}+m_{22}$$
Only being interested in the trace as a result, in a matrix multiplication
the order of factor matrices can be changed cyclically. As in (37b) every
{\tt bra} and every {\tt ket} as well as each operator part between the two
$\vert$ is a matrix we can rotate factors to obtain
\begin{eqnarray}
\sigma_c(\tau)
\left[ {\bf 1}\; \delta(\tau_n-\tau)+G^r(\tau,\tau_n)\sigma(\tau_n)\right]
\quad \vert\Psi^0(\tau_n)>\; \cdot\nonumber \\
<\Psi^0(\tau_0)\vert \Psi^0(\tau_0)>\frac{i}{\hbar}<\Psi^0(\tau_m)\vert \; 
\left[ {\bf 1}\; \delta(\tau_m-\tau)+\sigma(\tau_m)G^a(\tau_m,\tau)\right]
\end{eqnarray}
\renewcommand{\theequation}{\arabic{equation}}
\setcounter{equation}{39}
$<\Psi^0(\tau_0)\vert\Psi^0(\tau_0)>={\bf 1}$ again drops out.
$\vert\Psi^0(\tau_n)>\frac{i}{\hbar}<\Psi^0(\tau_m)\vert$ is of the structure
$g$, however, with no restriction as to which argument $\tau_n$ or $\tau_m$
is earlier or later in time. We define this new type of Green's function as
\begin{equation}
g^{+-}(\tau,\tau',\omega)=\frac{i}{\hbar}e^{-i\omega(\tau-\tau')}=
\begin{cases} -g^r(\tau,\tau',\omega) \quad {\rm if} \; \tau>\tau' \\
\phantom{-} g^a(\tau,\tau',\omega) \quad {\rm if} \; \tau<\tau' \end{cases}
\end{equation}
(An eventually ill-defined single point $\tau=\tau'$
is irrelevant for later integrations.)
As the cases $\tau>\tau'$ and $\tau<\tau'$ are mutually exclusive
the function can also be given by $g^{+-}(\tau,\tau')=g^a(\tau,\tau')-
g^r(\tau,\tau')$ (conclusions on $g^{+-}(\omega)$ from that are risky,
to my opinion, though). We have thus deduced the current formula
\begin{eqnarray}
I(\tau)&=&-e\; {\rm Tr}\; \{ \; \sigma_c(\tau) \int_{-\infty}^{\tau} d\tau_n
\int_{-\infty}^{\tau} d\tau_m \; \left[ ({\bf 1}\delta(\tau_n-\tau)+
G^r(\tau,\tau_n)\sigma(\tau_n)\right] \; \cdot\nonumber \\ && \cdot \;
g^{+-}(\tau_n,\tau_m) \; \left[ {\bf 1}\delta(\tau_m-\tau)+\sigma(\tau_m)
G^a(\tau_m,\tau) \right] \; \}
\end{eqnarray}
(like discussed in (20) it is rather irrelevant whether the upper integration
limits are set as $\tau$ or $\infty$)
or
\renewcommand{\theequation}{\arabic{equation}a}
\setcounter{equation}{40}
\begin{equation}
I(\tau)=-e\; {\rm Tr}\; \left\{ \sigma_c(\tau)\; G^{+-}(\tau,\tau) \right\}
\end{equation}
\renewcommand{\theequation}{\arabic{equation}}
\setcounter{equation}{41}
if $G^{+-}$ is defined as
\begin{eqnarray}
G^{+-}(\tau,\tau')=\int d\tau_1 \int d\tau_2 \; && [{\bf 1}\delta(\tau_1-\tau)
+G^r(\tau,\tau_1)\sigma(\tau_1)] \; g^{+-}(\tau_1,\tau_2) \cdot \nonumber \\
&& [{\bf 1}\delta(\tau_2-\tau)+\sigma(\tau_2)G^a(\tau_2,\tau')]
\end{eqnarray}
or
\renewcommand{\theequation}{\arabic{equation}a}
\setcounter{equation}{41}
\begin{equation}
G^{+-}=({\bf 1}+G^r\sigma)\; g^{+-}\; ({\bf 1}+\sigma G^a)
\end{equation}
\renewcommand{\theequation}{\arabic{equation}}
\setcounter{equation}{42}
in short notation. 
The current only needs $G^{+-}$ of two identical time arguments.
We shall need the Fourier representation of $g^{+-}(\omega)$, but none such for $G^{+-}$.
Although in our simple model for the normal-conducting contact the current
will come out the same for any $\tau$, $I(\tau)$ generally does depend on time
(our expectation value is an ensemble, not a time average). For the
superconducting case the ac parts \cite{Cuequ} like the Josephson current are included in
the expression (41), although one is mostly interested in the contribution
in $I$ that is independent of $\tau$ and gives the dc part.

Using $g^{+-}(\tau_n,\tau_m)$ without a third argument $\omega$ in (41)
we understood integration over frequency like in (16). But this point
requires more care as it has not been taken into account so far whether
states are occupied or empty at $\tau_0\rightarrow -\infty$. Writing out the
trace from (41a) in LR-components reveals four contributions
\begin{eqnarray}
-I/e &=& \sigma_{c,LR}\; (1+G^r_{RL}\sigma_{LR})\; g^{+-}_{RR} \;
\sigma_{RL}G^a_{LL}\nonumber \\
&+& \sigma_{c,LR} \; G^r_{RR}\sigma_{RL}\; g^{+-}_{LL}\;
(1+\sigma_{LR}G^a_{RL}) \nonumber \\
&+& \sigma_{c,RL}\; G^r_{LL}\sigma_{LR}\; g^{+-}_{RR}\;
(1+\sigma_{RL}G^a_{LR})\nonumber \\
&+& \sigma_{c,RL}\; (1+G^r_{LR}\sigma_{LR})\; g^{+-}_{LL}\;
\sigma_{LR}G^a_{RR}
\end{eqnarray}
It is clear that in contrast to the other appearing $\sigma$, in $\sigma_c$
from (36) there is an additional relativ minus sign between the LR- and the
RL-component. Obviously the net current is the difference between the
current from left to right and the current from right to left. In the first
- as well as the second - term in (43) a transition from R to L at $\tau$,
the time argument of $\sigma_c$, is picked out to be counted for the current
(see also Fig.4). With $g^{+-}$ at R, the charge carrier is supplied from a
state originally located on the right. It is an electron if the energy lies
below the right-side Fermi level. Until $\tau$ the state has evolved to
again be on the right. The originally left behind empty state at the same
energy below the Fermi level must have evolved to be on the left at $\tau$
such that by the R$\rightarrow$L transfer the electron can go into it
(Fig.5a). Or you might say that the plus and the minus branch of the Keldysh
contor represent two possible parts in the evolution of an original wave
function $\Psi^0_R$, between which there is a non-vanishing matrix element
of the operator $\sigma_c=[H,\rho_L]$. However, still regarding the first
line of (43), there is the further possibility that an empty state from
above the Fermi level on the right evolves to be at R at $\tau$ again, but
its left-behind complement (a negative charge) has evolved to be at L.
$\sigma_{LR}(\tau)$ in this case means the transition of an empty state or
positively charged particle from right to left (Fig.5b). Although one does not
usually introduce the concept of holes with transport in normal
conducting metals, it makes sense here to call unoccupied states simply
''holes''. In this way the model already includes the dual nature of charge
carriers needed for the superconducting case. In the normal conducting
case states do not change in nature (electron or hole) or energy during
their evolution (little zigzags are drawn in Fig.5 only to make the
multiple hoppings visible). At the superconducting junction, Andreev
reflection can be interpreted as changing an electron into a hole or vice
versa and mirroring its energy at the Fermi level \cite{my06} (see Fig.9 in section 6).
To have a charge carrier at a certain energy level at time $\tau$ to make a
certain transition, it is important that there was one at the corresponding
energy in the original uncoupled system. The coupling may have changed the
distribution compared to the occupation in uncoupled bulk reservoirs. And
the applied voltage imposes a non-equilibrium situation, anyway.
For the evolution of a state from the right as shown in Fig.5 it does not
matter whether the state at the respective energy on the left is occupied
or empty. If, for example the regarded energy lies below the Fermi levels
both left and right, there will be two states evolving as an electron on the
plus and a hole on the minus branch of the Keldysh contor, one having originated
at R and the other at L. These original, uncoupled and independent states are
our basis, especially for calculating an expectation value as trace. They
do not interfere. Terms with $g^{+-}_{RR}$ and $g^{+-}_{LL}$ are simply
added in (43). Schemes analogous to Fig.5 could be drawn for the terms from
the last three lines of (43) as well. The conclusion of the whole argumentation
of how to let the original Fermi occupation function for the reservoirs left
and right enter the current calculation is that $g^{+-}_{LL}$ has to change
sign at the left Fermi energy and $g^{+-}_{RR}$ at the right Fermi energy.
Let us note $g^{+-}$ like $g^r$ and $g^a$ for any bulk reservoir with Fermi
level at $\omega=0$. We shall keep $g^{+-}$ corresponding to $g^r$ and $g^a$
as in (40) for occupied electron states below the Fermi level and change the
sign for empty states above it.
\begin{equation}
g^{+-}(\tau,\tau')=\frac{\mathcal{D}}{h}\int d\omega \; g^{+-}(\omega) \;
e^{-i\omega(\tau-\tau')} \; {\rm with} \; g^{+-}(\omega)=\begin{cases}
\phantom{-}i \; {\rm for}\; \omega<0 \\ -i \; {\rm for}\; \omega>0\end{cases}
\end{equation}
Although it might be practical to use  (19), (23), (44), (18) and (36) in (41)
to quite directly produce an expression that calculates the current finally
as an integral over frequency and in our simple model can even be analytically
evaluated, in parallel to \cite{Cuequ} we shall use the transfer Green's functions
here. The current is best translated into an expression of the transfer
functions from the form already resolved into LR-components (43). Furthermore
eliminate $\sigma_c$ through $\sigma_{c,LR}=-\sigma_{LR}$ and
$\sigma_{c,RL}=\sigma_{RL}$. L- and R-indices follow logically from (26), for
example $G_{LL}\sigma_{LR}=g_{LL}T_{LR}$.
\renewcommand{\theequation}{43\alph{equation}}
\setcounter{equation}{0}
\begin{eqnarray}
I/e&=&\sigma_{LR}\; (1+g^r_{RR}T^r_{RR}) \; g^{+-}_{RR} \; T^a_{RL}g^a_{LL}
\nonumber \\
&+& \sigma_{LR} \; g^r_{RR}T^r_{RL} \; g^{+-}_{LL} \; (1+T^a_{LL}g^a_{LL})
\nonumber \\
&-& \sigma_{RL} \; g^r_{LL}T^r_{LR} \; g^{+-}_{RR} \; (1+T^a_{RR}g^a_{RR})
\nonumber \\
&-& \sigma_{RL} \; (1+g^r_{LL}T^r_{LL}) \; g^{+-}_{LL} \; T^a_{LR}g^a_{RR}  
\end{eqnarray}
As quantities here are no longer matrices, but simply functions of time or
frequency, the leading $\sigma_{LR}$ and $\sigma_{RL}$ in the second and third
line of (43a) can be moved to the ends of the products. From (41) we remember
that their argument is $\tau$, the same as the second argument of the last
factor to the right. Then using relations from (27a) and complementary forms
the current formula simplifies to
\begin{eqnarray}
I/e&=& T^r_{LR} g^{+-}_{RR} T^a_{RL}g^a_{LL} \; + \;
g^r_{RR} T^r_{RL} g^{+-}_{LL} T^a_{LR} \nonumber \\
&-& g^r_{LL} T^r_{LR} g^{+-}_{RR} T^a_{RL} \; - \;
T^r_{RL} g^{+-}_{LL} T^a_{LR} g^a_{RR}
\end{eqnarray}
\renewcommand{\theequation}{\arabic{equation}}
\setcounter{equation}{44}
The terms with the factors 1 from the brackets have elegantly been made to
vanish. Now we use the Fourier representations for all functions in (43b).
As all terms follow the same scheme, the second is treated in an exemplary way
here (all integrals run from $-\infty$ to $+\infty$):

\begin{eqnarray}
\int d\tau_1\; d\tau_2\; d\tau_3\; d\omega_1\; d\omega\; d\omega_2\; d\omega_3\;
\frac{\mathcal{D}}{h} g^r_{RR}(\omega_1) e^{-i\omega_1\tau} e^{i\omega_1\tau_1}
\frac{\hbar}{2\pi\mathcal{D}}T^r_{RL}(\omega_2)\nonumber \\
e^{-i(\omega_2+eV/\hbar)\tau_1} e^{i\omega_2\tau_2} \;
\frac{\mathcal{D}}{h} g^{+-}_{LL}(\omega) e^{-i\omega\tau_2} e^{i\omega\tau_3} 
\frac{\hbar}{2\pi\mathcal{D}} T^a_{LR}(\omega_3)
e^{-i\omega_3\tau_3} e^{i(\omega_3+eV/\hbar)\tau} \nonumber
\end{eqnarray}
Doing the $\tau_1$,$\tau_2$- and $\tau_3$-integrals gives $(2\pi)^3\delta(
\omega_1-\omega_2-eV/\hbar)\delta(\omega_2-\omega)\delta(\omega-\omega_3)$.
Then even the exponentials with $\tau$ cancel and the term simplifies to
$$ \frac{1}{2\pi}\int d\omega\; g^r_{RR}(\omega+eV/\hbar)\; T^r_{RL}(\omega)\;
g^{+-}_{LL}(\omega)\; T^a_{LR}(\omega) $$
In our case (31) tells us that $T^{r/a}_{LR/RL}(\omega)$ are all identical and
real, $g^r_{LL/RR}(\omega)=-i$ is the complex conjugate of $g^a_{LL/RR}(\omega)
=i$ and $-g^{+-}_{LL/RR}$ is the complex conjugate of
$g^{+-}_{LL/RR}$ of the same $\omega$ as $g^{+-}(\omega)$ is purely imaginary,
too. Thus the last two terms in (43b) are the complex conjugates of the first
two and thus twice the real part of these first two can be taken for $I/e$.
In the superconducting version of the model, where $T$ and $g$ actually are
$\omega$-dependent, complex conjugate relations \cite{my06} also exist between
$T$s as well as $g$s, and the current formula can be reduced in the same way.
Just to note the quite general formula \cite{Cuequ} in short form:
\begin{equation}
I/e=2\; {\rm Re}\; \left\{  T^r_{LR} g^{+-}_{RR} T^a_{RL} g^a_{LL} +
g^r_{RR} T^r_{RL} g^{+-}_{LL} T^a_{LR} \right\}
\end{equation}
For the integrand from the second term in our model we get
$$ g^r_{RR}(\omega+eV/\hbar) T^r_{RL}(\omega) g^{+-}_{LL}(\omega) T^a_{LR}(
\omega)=(-i) \frac{t}{1+t^2} (\mp i) \frac{t}{1+t^2} =
\mp \frac{t^2}{(1+t^2)^2} $$
where the signs refer to $\omega$ greater or less than zero and come out reversed
for the first term with $g^{+-}_{RR}$, because there is $g^a$ instead of $g^r$.
Care has to be taken with the reference point for $\omega$ in both terms.
This may easily be overlooked in the normal conducting case here in contrast
to the superconducting case where $T$ indeed is $\omega$-dependent and like
$G$ as a function of only one argument always referred to the same Fermi
level (the left, for example). If we call the argument of $g^{+-}_{LL}$
from the second term $\omega$, the one for $g^{+-}_{RR}$ in the first term
is $\omega+eV/\hbar$. On an $\omega_{(L)}$-axis, the second term
changes sign at zero, however, the first jumps at $-eV/\hbar$ (Fig.6). A shift
of the integration parameter cannot be made independently for both terms.
Thus, for the normal conducting model here
\renewcommand{\theequation}{45\alph{equation}}
\setcounter{equation}{0}
\begin{eqnarray}
I/e=2\; {\rm Re} \; \frac{1}{2\pi} \int d\omega \; \{ \; T^r_{LR}(\omega)
g^{+-}_{RR}(\omega+\frac{eV}{\hbar}) T^a_{RL}(\omega) g^a_{LL}(\omega)
\nonumber \\ + \;
g^r_{RR}(\omega+\frac{eV}{\hbar}) T^r_{RL}(\omega) g^{+-}_{LL}(\omega)
T^a_{LR}(\omega) \; \} \end{eqnarray}
\renewcommand{\theequation}{\arabic{equation}}
\setcounter{equation}{45}
In principle the convention is needed, that the $T$-argument always refers
to the left side, but for $g$ general formula like (44) for a single bulk
with Fermi level at zero frequency are applied. A more involved situation
where choosing integration intervals consistently for all contributing current
terms is crucial can be found in \cite{my06}.

\begin{figure} \begin{center}
\includegraphics[width=12cm]{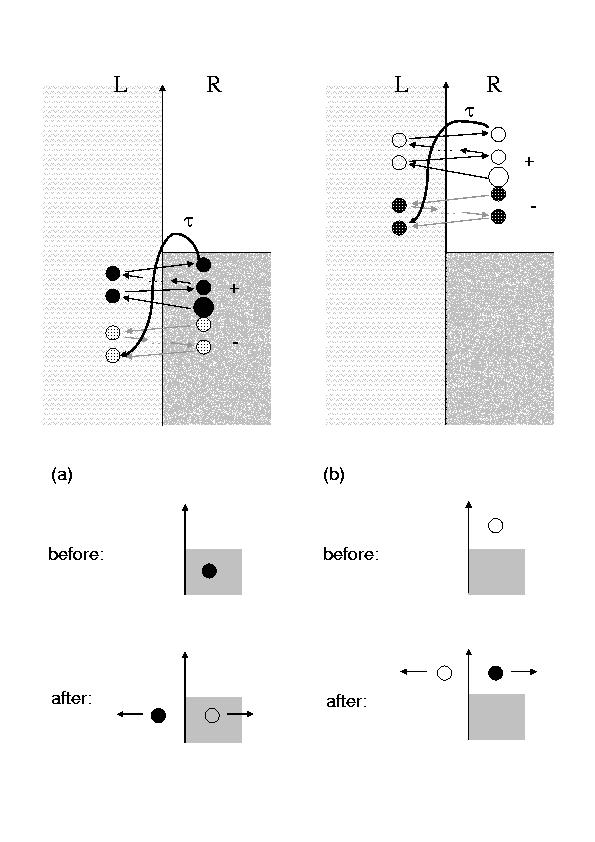}
\end{center}
\caption{(a) Evolution of an electron state from R such that at $\tau$ the
charge is transferred to the left. (b) Evolution of an unoccupied state above
the Fermi level from R such that at $\tau$ it gets filled by a charge from
the left or the hole state shifts to the left. The lower pictures sketch the
effect of the original uncoupled state for the current. Sending electrons
from the right to the left will be compensated by electrons sent from left
to right. But with levels as in Fig.1c for holes sent from right to left there
will be a range between the two Fermi levels where there are no counterbalancing
holes going the other way. (For the left, electrons going to the right are not
outweighed in this range.)}
\end{figure}

\begin{figure} \begin{center}
\includegraphics[width=8cm,angle=270]{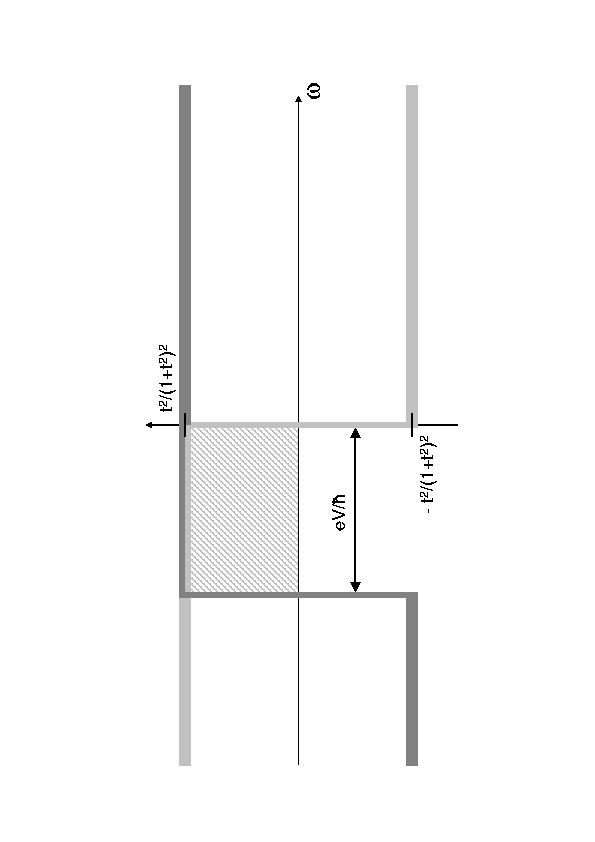} \end{center}
\caption{The two terms under the integral from equation (45). They cancel
except on an interval of length $eV/\hbar$.}
\end{figure}

From Fig.6 it is easily seen that that the integral is twice the constant
$\frac{t^2}{(1+t^2)^2}$ integrated over an interval of length $eV/\hbar$,
and outside that interval contributions cancel. With the other factor 2 from
twice the real part and the prefactor $1/2\pi$
the result for the current finally is
\begin{eqnarray} &&
I/e=2\cdot \frac{1}{2\pi}\cdot 2\cdot \frac{t^2}{(1+t^2)^2}\cdot\frac{eV}{\hbar}
\quad\quad \Longleftrightarrow \nonumber \\ &&
\frac{V}{I}=\frac{h}{e^2}\left(\frac{4t^2}{(1+t^2)^2}\right)^{-1}
\qquad \Longleftrightarrow \qquad
\frac{I}{V}=\frac{e^2}{h} \frac{4t^2}{(1+t^2)^2}
\end{eqnarray}
The factor $\theta=4t^2/(1+t^2)^2$ is the conductance in units of the quantum
conductance or its inverse the resistance in units of the quantum resistance
\cite{Cuequ}.
(The conductance of a channel doubles if two spin states are allowed. Then
$h/2e^2=13k\Omega$ should be taken as resistance unit.)
$\theta$ is the transmission probability of the conductance channel through
the junction we regarded. The result that in the normal-conducting case
the current is proportional to the voltage is not at all surprising, of course.
The non-trivial result is the conversion of the quantum mechanical transmission
amplitude $t$ to the measurable transmission probability $\theta$.
$\theta=0$ if $t=0$. And a totally open channel with $t=1$ has
transmission probability $\theta=1$. 

It may seem a contradiction on the one hand calculating the current from a
changing amount of charge on one side and on the other hand saying that missing
charges are replaced and superfluous ones led away by the voltage source.
A slightly different viewpoint may help to get convinced that the calculated
quantity is indeed the current in the stationary, but non-equilibrium system.
The crucial point was putting on to evaluate $d\rho_L/dt$ in state $\Psi(t)$.
Without further ado we could not tell whether this state of the coupled system
was occupied or not. The Schr\"odinger equation (5) set up the left and right
material properties as well as the coupling across the junction, however,
did not take any account of the effect of the voltage source. Without need
to specify real locations for the division, just principally view our structure
as consisting of a junction region and leads. The $\Psi(t)$ defined through (5)
describes states in the junction region. But think of them as offered by the
system and following their time development whether occupied or not. The leads
are always occupied exactly up to their Fermi levels. Electrons freshly
supplied by the voltage source need not be in phase with present ones. A
random phase is most easily modelled by assuming the left lead wave function
at any time $\begin{pmatrix} 1 \cr 0\cr\end{pmatrix}$ without a phase in contrast to the
assumed $\begin{pmatrix} \exp(-i\omega_L\tau)\cr 0\cr\end{pmatrix}$ for the left side of the
junction region. Which states to which extent actually get occupied in the
left side of the junction region, that is $\rho_L(t)$, is determined by the
overlap of $\Psi(t)$ with the phaseless left lead wave function
$\begin{pmatrix} 1\cr 0\cr\end{pmatrix}$,
which we recognize as $\Psi_L^0(-\infty)$.\ $\rho_L=\vert <
\begin{pmatrix}1\cr 0\cr\end{pmatrix}\vert\begin{pmatrix}\Psi_L(t)\cr 0\cr\end{pmatrix}>\vert^2$. The current
then is the change in time of $\rho_L$ due to charge flow through the junction,
that is processes inside the junction region only, corresponding to what our
Hamiltonian was set up for. A similar line of thought with state overlaps
can be applied to the passing of charges out of the junction region into
the (right) lead; it may be helpful to view this as putting empty states or
holes into the junction region, though. Of course, only the L-part of $\Psi(t)$
can overlap with $\begin{pmatrix} 1\cr 0\cr\end{pmatrix}$. $\begin{pmatrix}
\Psi_L(t)\cr 0\cr\end{pmatrix}$ can thus
be replaced by $\Psi(t)$ in our new expression for $\rho_L$. Recalling the
operator definition (34) the time derivative of $\rho_L$
$$
\frac{d}{dt}\; \rho_L=\frac{d}{dt}\; \vert < \begin{pmatrix}1\cr 0\cr\end{pmatrix}\vert \Psi(t)>
\vert^2=\frac{d}{dt} \left\{ <\Psi(t) \vert \begin{pmatrix}1\cr 0\cr\end{pmatrix}><
\begin{pmatrix} 1\cr 0\cr\end{pmatrix} \vert \Psi(t) > \right\}
$$
is identical to the ansatz made by putting together (33) and (34) in (37).

\begin{figure}\begin{center}
\includegraphics[width=8cm,angle=270]{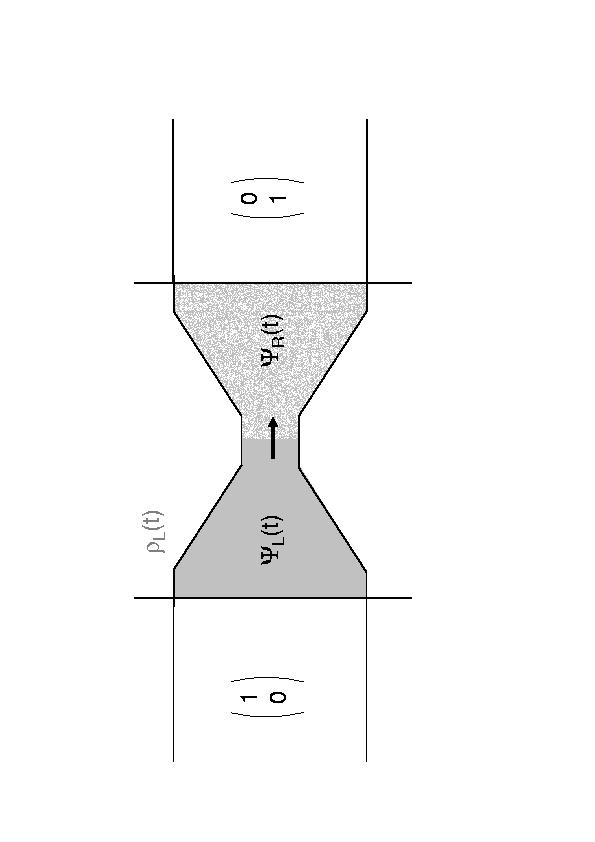}
\end{center}
\caption{Hypothetical division of the structure into the actual junction
region and bulk leads.}
\end{figure}

\subsection{Comment on the two-level system}
At first glance the problem posed by the Schr\"odinger equations without and
with coupling, (1) and (5), especially if we regard a single energy level on
each side with corresponding $\omega_L$ and $\omega_R$ as in Fig.1c, looks
like the two-level system known from quantum mechanics textbooks \cite{CoTan}.
By the coupling the two energy levels are shifted. The new eigenstates lie
further apart from one another. If the system is initiated in one of the
uncoupled states, it will oscillate harmonically between the two levels,
and both both the period of the oscillation as well as the maximum
transition probability to the other state depend on the energy difference of
the original states.

\begin{figure}\begin{center}
\includegraphics[width=7cm]{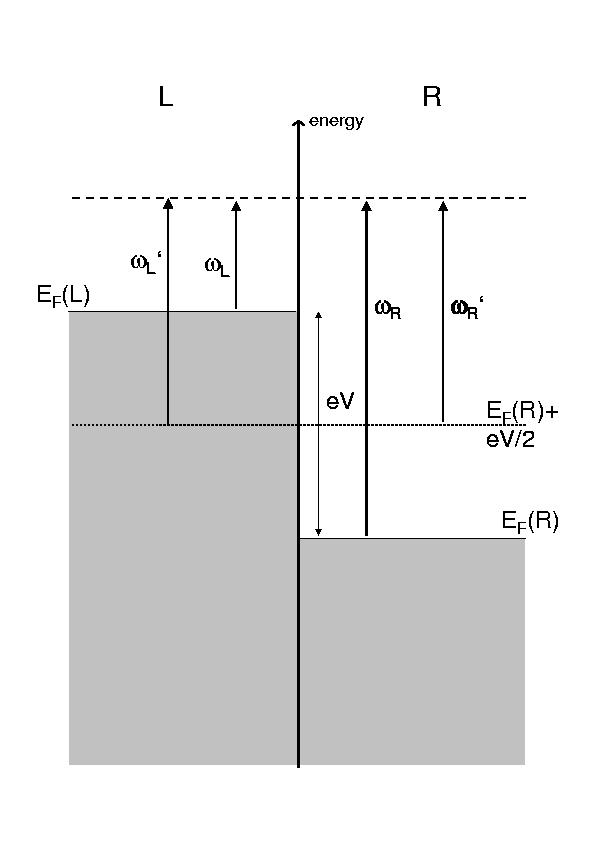}
\end{center}
\caption{Attempt to adjust energy levels from left and right to same
reference.}
\end{figure}

Even if the time dependence of the interaction (18) can be got rid of by
changing to the interaction picture, it makes no sense to numbly calculate
eigenvalues and eigenvectors of a Hamiltonian $\begin{pmatrix} \hbar\omega_L & W\cr
W & \hbar\omega_R\cr\end{pmatrix}$. It is not clear from which reference levels such
eigenenergies $E_{\pm}$ should be counted as zero levels are different for
$\omega_L$ and $\omega_R$. If one tries to refer the left and right energies
already of the uncoupled system to the same reference level, $E_F(R)+\frac{eV}{2}$
for example (Fig.8), $\Psi^0_L$ (and $\Psi^0_R$) cannot be taken as eigenstates any
more, because from $i\hbar\partial_{\tau}e^{-i\omega_L\tau}=\hbar\omega_L
e^{-i\omega_L\tau}$ with $\omega_L=\omega_L'+\frac{eV}{2}$ it does {\it not}
follow that $i\hbar\partial_{\tau}e^{-i(\omega_L'+eV/2)\tau}$ equals
$\hbar\omega_L'e^{-i(\omega_L'+eV/2)\tau}$. The standard treatment of the
two-level system is not applicable to the non-equilibrium situation.

\subsection{Superconducting junction}
Despite having been the simplest example to introduce the Green's functions
scheme, applying the formalism to the normal conducting junction to get out that
the current is proportional to the applied voltage was breaking a butterfly
on the wheel. Although repeatedly mentioned, fully developing the extension
to the superconducting case is beyond the scope of this presentation. But the
resulting current-voltage characteristics shall be shown as a plea for the
usefulness of the method. A different approach based on matching wave
functions \cite{AveBar} leads to identical results, though.

\begin{figure} \begin{center}
\includegraphics[width=8cm,angle=270]{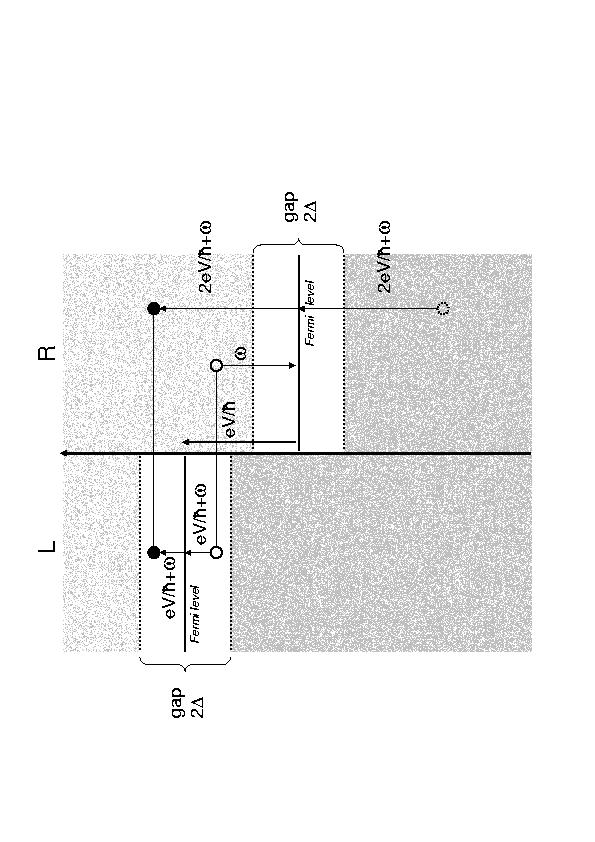}\end{center}
\caption{Illustration of Andreev reflection as mirroring at the Fermi levels
for electron-hole conversion and vice versa. For either side the incoming and
reflected levels can lie inside or outside the gap. Only the beginning and end
of a complete multiple-reflection process have to be in the electron or hole
reservoir below or above the gap, respectively.}
\end{figure}

Some formula shall be listed, because they are not necessarily written out
in complete analogy to the presentation here in \cite{Cuequ} and other literature.
Working in the quasiparticle picture, in the superconducting case each entry
in LR-space of a Green's or transfer function expands into another 2x2 matrix
in Nambu space over electrons and holes. There is \cite{ketterson_2}:
\begin{equation}
g^{r/a}_{LL/RR}(\omega)=\begin{pmatrix} g^{ee} & g^{eh} \cr g^{he} & g^{hh} \cr\end{pmatrix}^{
r/a}_{LL/RR}(\omega)=\frac{1}{\sqrt{(\Delta/\hbar)^2-(\omega\pm i\eta)^2}}
\begin{pmatrix} -\omega\mp i\eta & \Delta/\hbar \cr -\Delta/\hbar & \omega\pm i\eta\cr\end{pmatrix}
\end{equation}
$g$ is the same for LL and RR. Different signs refer to the retarded and advanced
function. $\Delta$ is half the gap of the superconductor. The small imaginary
part $\eta$ is added to get the correct root besides slightly smoothening
singularities. $g(\tau,\tau') =\mathcal{D}/h\; \int d\omega \; g(\omega)\;
e^{-i\omega\tau} \; e^{i\omega\tau'}$ still holds.
\begin{equation}
\sigma_{LR/RL}=\begin{pmatrix} \sigma^{ee} & 0 \cr 0 & \sigma^{hh}\end{pmatrix}_{LR/RL}=
\frac{\hbar}{\mathcal{D}}\;
\begin{pmatrix} t\; e^{\pm i eV\tau/\hbar} & 0 \cr 0 & -t\; e^{\mp ieV\tau/\hbar}\end{pmatrix}
\end{equation}
Here signs refer to the LR- and RL-direction, respectively. Remark the
reversed signs for holes in the exponential compared to electrons.
There is no electron-hole conversion during a single hopping. The $eh$-
and $he$-elements of $g$ introduce so-called Andreev reflection.
(Cooper-pair tunneling \cite{Urbina} is not included.) $T$ is a full matrix in $eh$-space
and has to be set up as
\begin{equation}
\frac{\mathcal{D}}{\hbar} \; T(\tau,\tau') =
\frac{1}{2\pi} \sum_n \int d\omega \; T_n(\omega)\;
e^{-i\omega\tau} \; e^{ineV\tau/\hbar}\; e^{i\omega\tau'}
\end{equation}
We skipped the distinction of reference $\omega_{J/K}$ as in (25), because in
practice only one type, e.g. $T_{LR}$, has to be calculated, as then $T_{RL}$
follows from complex conjugation. $n$ runs over all integers. A Dyson equation
like (32) leads to a recursion which connects $T_n$ to $T_{n-2}$ and $T_{n+2}$.
It can be solved by reasonably truncating the $n$-range. Then, just also
summing over $n$, the established current formula (45) can be used.

Expressions like (32) have to sum over all possible combinations of $e$- and
$h$-indices. For example:
\begin{eqnarray}
T^{ee}_{LR}=\sigma_{LR}^{ee}&+&
\sigma_{LR}^{ee}g^{ee}_{RR}\sigma_{RL}^{ee}g^{ee}_{LL}T^{ee}_{LR}+
\sigma_{LR}^{ee}g^{eh}_{RR}\sigma_{RL}^{hh}g^{he}_{LL}T^{ee}_{LR}\nonumber \\ &+&
\sigma_{LR}^{ee}g^{ee}_{RR}\sigma_{RL}^{ee}g^{eh}_{LL}T^{he}_{LR}+
\sigma_{LR}^{ee}g^{eh}_{RR}\sigma_{RL}^{hh}g^{hh}_{LL}T^{he}_{LR}
\end{eqnarray}

\begin{figure}\begin{center}                            
\includegraphics[width=7cm,angle=270]{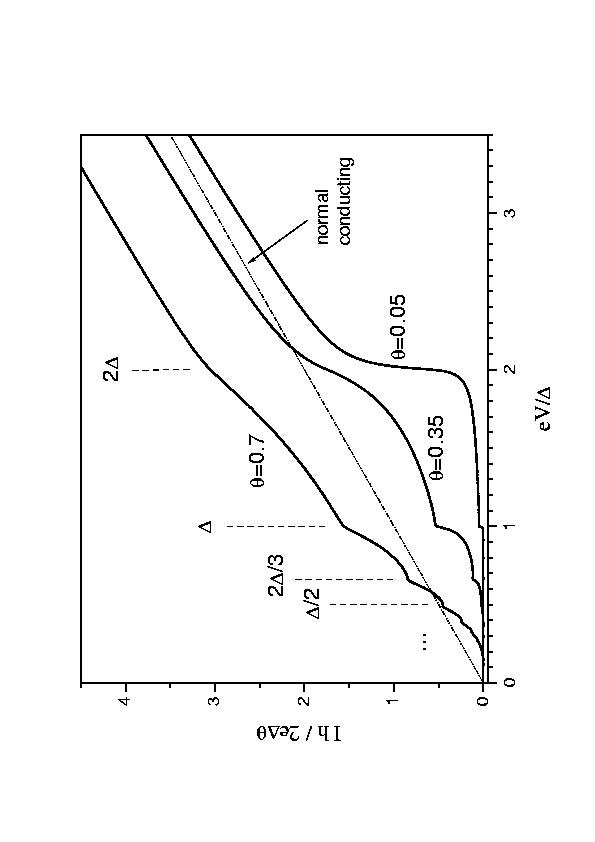}\end{center}
\caption{Calculated current-voltage curves for transport channels of three
different transmissions (normalized to $\theta$) [reproduced after \cite{Cuequ}].
The normal conducting $IV$ is added for comparison (keep $\Delta$ just for units).}
\end{figure}

To point out the Andreev reflection in the model, we look at a term
$g\sigma g\sigma g$ with alternating $e$- and $h$-indices from the
development (21). Such a sequence will also be contained as parts in
higher-order terms from (21) or (24). Contracting inner time arguments
results in
\begin{eqnarray} &&
\int d\tau_1\; d\tau_2\; g^{(h)e}_{RR}(\tau,\tau_1) \; \sigma^{ee}_{RL}(
\tau_1)\; g^{eh}_{LL}(\tau_1,\tau_2)\; \sigma^{hh}_{LR}(\tau_2) \;
g^{h(e)}_{RR}(\tau_2,\tau') \nonumber \\ &=&
\int d\tau_1\; d\tau_2\; d\omega_2 \; d\omega_1\; d\omega\;
\frac{\mathcal{D}}{h} g^{(h)e}_{RR}(\omega_2)e^{-i\omega_2\tau}e^{i\omega_2\tau_1}\;
\frac{\hbar}{\mathcal{D}} t e^{-ieV\tau_1/\hbar} \; \cdot
\nonumber \\ && \phantom{++}
\frac{\mathcal{D}}{h} g^{eh}_{LL}(\omega_1)e^{-i\omega_1\tau_1}e^{i\omega_1\tau_2}\;
\frac{\hbar}{\mathcal{D}} (-t) e^{-ieV\tau_2/\hbar} \;
\frac{\mathcal{D}}{h} g^{h(e)}_{RR}(\omega) e^{-i\omega\tau_2}e^{i\omega\tau'}
\nonumber \\ &=& (-t^2)\; \frac{\mathcal{D}}{h} \int d\omega \; d\omega_1\; d\omega_2\; 
\delta(\omega_2-\frac{eV}{\hbar}-\omega_1) \;
\delta(\omega_1-\frac{eV}{\hbar}-\omega) \; \cdot \nonumber \\ && \phantom{+++++}
g^{(h)e}_{RR}(\omega_2) \; g^{eh}_{LL}(\omega_1) \; g^{h(e)}_{RR}(\omega) \;
e^{-i\omega_2\tau}\; e^{i\omega\tau'} \nonumber  \\ &=&
-t^2\; \frac{\mathcal{D}}{h}\int d\omega\; g^{(h)e}_{RR}(\omega+2\frac{eV}{
\hbar})\; g^{eh}_{LL}(\omega+\frac{eV}{\hbar}) \; g^{h(e)}_{RR}(\omega)\;
e^{-i(\omega+2eV/\hbar)\tau}\; e^{i\omega\tau'} \nonumber \\ && \phantom{Zeile}
\end{eqnarray}
because the $\tau_1$ and $\tau_2$-integration give $4\pi^2\delta(\omega_2-
eV/\hbar-\omega_1)\delta(\omega_1-eV/\hbar-\omega)$.
In Fig.9 for holes the energy axis has to be reversed as compared to electrons.
Therefore the initial $\omega$ is negative. For the argument here, it does
not matter if the beginning factor in (51) is $g^{he}_{RR}$ or $g^{ee}_{RR}$.
If it were $h$, then the Andreev reflection marked by the dashed arrow would
follow. The last index of the last $g_{RR}$ could also be $h$ instead of $e$. 
Whereas in the normal conducting case in multiple reflections the charge
carrying particle always came back to the same energy with regard to the
Fermi level when coming back to the same side of the junction (Fig.1), in
(51) the frequency argument of the first $g_{RR}$ is shifted by twice the
applied voltage equivalent with respect to the second $g_{RR}$
(Fig.9). Such shifts enable transport even when the gaps of the left and
right side are still overlapping, that is for voltages below $2\Delta/e$
(Fig.10). In the $IV$-curves steps towards lower voltages at fractions of
$2\Delta$ are associated to ever higher-order Andreev reflections. Of course,
these current contributions
become the more prominent the greater the channel transmission.
Besides the step heights and positions the curvature on each step is an
important signature. The model agrees excellently with experimental results
\cite{Sche97}.
The interference of transport processes of different order is correctly
taken into account. Simpler models \cite{Bratus} that add up tranfer probabilities
proportional to $\theta^{order}$ cannot reproduce these characteristics and
are only valid for low transmissions $\theta<<1$.

\section{Conclusions and Outlook for section III}
It has been explained on a quite basic level how quantum transport between
two reservoirs in a stationary non-equilibrium state can be modelled. The
purpose was to present a Green's functions technique for handling coupling
in the context of a field of current research interest, namely transport
through point contacts. Although besides the general formalism (also see section I)
requiring the development of quite some more subject specific mathematical
framework such as the transfer Green's functions, the transport through a
contact with arbitrary transmission is a suitable example to  illustrate the
inclusion of interaction up to all orders in the implicit Dyson equations.
Calculations
have been carried through in every detail for the normal-conducting single
junction. Basic formula and results have been given for the superconducting
junction. The decomposition into transport channels (eigenmodes) of a point
contact can be inferred from its superconducting transport characteristics
which can be taken like a PIN-code \cite{Sche97}. The presented Green's functions formalism
has a great potential for extension. Systems of two more or less coherently
linked junctions can be modelled \cite{my06}.
Models for transport  through molecules \cite{Heurich,Roland,Wohltat} or
atomic chains \cite{Mertig} so far mainly rely on ab initio calculations of
the density of states. Time dependent density functional theory for
non-equilibrium situations is also developed  \cite{Kurth}.

\begin{acknowledgements}
The author thanks Alain Dereux for learning about Green's functions in
electrodynamics, Juan-Carlos Cuevas for through his former work introducing
her to Green's functions in quantum transport, her supervisor Elke Scheer
for the freedom to work on this subject as well as Klaus-Ulrich Neumann
for the suggestion to write a paper of this kind starting from some
lecture notes.
\end{acknowledgements}

\end{document}